\documentclass[a4paper,12pt]{article}
\usepackage{epsfig}
\date{\today}
\newcommand{\be}{\begin{equation}}
\newcommand{\ee}{\end{equation}}
\newcommand{\ba}{\begin{eqnarray}}
\newcommand{\ea}{\end{eqnarray}}

\def\lsim{\raise0.3ex\hbox{$\;<$\kern-0.75em\raise-1.1ex\hbox{$\sim\;$}}}
\def\gsim{\raise0.3ex\hbox{$\;>$\kern-0.75em\raise-1.1ex\hbox{$\sim\;$}}}

\def\be{\beta}

\def\nn{\nonumber}

\begin{document}
\rightline{SUSX-TH/01-016}
\vskip 1cm
\begin{center}
{\bf \large{ EDM Constraints in Supersymmetric Theories\\[10mm]}}
{S. Abel$^{1,2}$, S. Khalil$^{1,3}$, and O. Lebedev$^1$ \\[6mm]}
\small{$^1$Centre for Theoretical Physics, University of Sussex, Brighton BN1
9QJ,~~U.~K.\\[4mm]}
\small{$^2$IPPP, University of Durham, South Rd., Durham DH1 3LE
,~~U.~K.\\[4mm]}
\small{$^3$Ain Shams University, Faculty of Science, Cairo, 11566, Egypt.\\[7mm]}
\end{center}

\vskip 0.3cm
\begin{minipage}[h]{14.0cm}
\begin{center}
\small{\bf Abstract}\\[3mm]
\end{center}
We systematically analyze  constraints on supersymmetric theories imposed by the 
experimental bounds on the electron, neutron, and mercury electric dipole moments.
We critically reappraise the known mechanisms to suppress the EDMs and conclude
that only the scenarios with  approximate CP-symmetry or flavour-off-diagonal
CP violation remain attractive after the addition of the mercury EDM  constraint.
\end{minipage}
\vskip 0.2cm
\vskip 1cm

\section{Introduction}

There are a number of reasons to suspect that 
there are additional sources of CP violation beyond those of the 
Standard Model (given by $\bar\theta$ and $\delta_{KM}$).
The most compelling one  is that  the SM is unable to explain the 
cosmological baryon asymmetry  of our universe. 
Also, the Standard Model  is very unlikely to be the ``ultimate'' theory
of nature. Most extensions of the SM bring in new sources of CP-violation.
In particular, the most attractive one  -- the Minimal
Supersymmetric Standard Model (MSSM) allows for  new sources
of CP violation in both  supersymmetry-breaking and supersymmetry-conserving
sectors, and there are no compelling arguments for them to be zero.
In addition, the improving precision in the measurements of the CP-observables 
such as  $A_{CP}(B\rightarrow \psi K_s)$ \cite{babar} may soon reveal deviations
from the SM predictions.

The most stringent constraints on models with additional sources of CP violation come 
from continued efforts to measure the electric 
dipole moments (EDM) of the neutron~\cite{bound},  electron~\cite{eedm},
and mercury atom \cite{mercury}
\begin{eqnarray} 
d_n &<& 6.3 \times 10^{-26} ~\mathrm{e~cm}~(90\% CL), \nonumber\\
d_e &<& 4.3 \times 10^{-27} ~\mathrm{e~cm}~,\nonumber\\
d_{Hg} &<& 2.1 \times 10^{-28} ~\mathrm{e~cm}~.
\end{eqnarray}
With the expected improvements in experimental precision,  
the EDM is likely to be one of the most important tests for
physics beyond the Standard Model for some time to come, and EDMs will remain a difficult hurdle for 
supersymmetric theories if they are to allow sufficient baryogenesis. 
Indeed it is remarkable that the SM contribution to the EDM of the
neutron is of order $10^{-30}$ e cm, whereas the ``generic'' 
supersymmetric value is $10^{-22}$e cm. 

In this paper we analyze neutron, electron, and mercury 
EDMs in the context of R-parity conserving supersymmetric theories. 
In particular, we reconsider the known mechanisms to suppress EDMs in light of
the recently reported bound on the mercury EDM  \cite{mercury}.
These include  SUSY models with small CP phases, models with heavy sfermions,
the cancellation scenario, and models with flavour off-diagonal CP violation.
We also study to what extent different scenarios rely on assumptions about the neutron
 structure, i.e. chiral quark model {\em vs} parton model. 

The paper is organized as follows. In section 2 we present general formulae for the
EDMs and discuss their model-dependence. In section 3 we define our
supersymmetric framework and present all relevant supersymmetric contributions 
to the EDMs. In particular, we analyze the importance of the two-loop Barr-Zee
and Weinberg type EDM contributions. Section 4 is devoted to the study of the
EDM suppression mechanisms. First we consider in detail  the ``canonical'' scenarios:
suppression due to small CP phases or heavy sfermions. Second, in the context
of the cancellation scenario, we analyze the possibility of the EDM cancellations
in two classes of models: mSUGRA-like models with nontrivial gaugino phases and
D-brane models. Finally, we discuss the EDM suppression in models with flavour-off-diagonal
CP violation. In addition, we present new model-independent bounds on the sfermion
mass insertions imposed by the electron, neutron, and mercury electric dipole moments.
In section 5 we overview and discuss our results.

\section{Electron, neutron, and mercury electric dipole moments}

Let us first summarize the contributions to the three most  
significant EDMs, beginning with the most reliable, the electron EDM. 

\subsection{Electron EDM}

The electron EDM is defined by the effective CP-violating interaction
\begin{equation}
{\cal{L}}= -{i\over 2} d_e \bar e \sigma_{\mu\nu}\gamma_5 e \;F^{\mu\nu}\;,
\end{equation}
where $F^{\mu\nu}$ is the electromagnetic field strength.
The experimental bound on the electron EDM is derived from the electric dipole
moment of the thallium atom and is given by \cite{eedm}
\begin{equation}
d_e < 4 \times 10^{-27} {\rm e\;  cm}\;.
\end{equation}
In supersymmetric models, the electron EDM arises due to CP-violating 1-loop diagrams with
the chargino and neutralino exchange:
\begin{equation}
d_e=d_e^{\chi^+}+d_e^{\chi^0}\;.
\end{equation}
Since the EEDM calculation involves little uncertainty it allows to extract reliable
bounds on the CP-violating SUSY phases.

\subsection{Neutron EDM}
The neutron EDM has contributions from a number of CP-violating operators involving quarks,  
gluons, and photons. The most important ones include the electric and chromoelectric
dipole operators, and the Weinberg three-gluon operator:
\begin{eqnarray}
 {\cal{L}}=&-& {i\over 2} d_q^E \bar q \sigma_{\mu\nu}\gamma_5 q \;F^{\mu\nu}\;
- \; {i\over 2} d_q^C \bar q \sigma_{\mu\nu}\gamma_5 T^a q \;G^{a \mu\nu }\; \nonumber\\
&-& \; {1\over 6} d^G f_{abc} G_{a \mu \rho} G_{b \nu}^{\rho}
G_{c \lambda \sigma} \epsilon^{\mu\nu\lambda\sigma}\;,
\end{eqnarray} 
where $G_{a \mu\nu} $ is the gluon field strength, $T^a$ and $f_{abc}$ are the
SU(3) generators and group structure coefficients, respectively.
Given these operators, it is however a nontrivial task to evaluate the neutron EDM
since  assumptions about the neutron internal structure are necessary. In what follows
we will study two models, namely the quark chiral model and the quark parton model.
Neither of these models is sufficiently reliable by itself \cite{pospelov}, 
however a power of the combined analysis should provide an insight into implications
of the bound on the neutron EDM and in particular comparing them gives some 
indication of the importance of these systematic errors in, for example, 
cancellations. A better justified approach to the neutron EDM based on
the QCD sum rules  has  appeared in \cite{pospelov1} and earlier work
\cite{khrip1}, \cite{khrip}. We note that in any case the NEDM calculations involve uncertain hadronic parameters 
 such as the quark masses and thus these calculations have a status of estimates. 
The major conclusions of the present work are independent of the specifics of the neutron model.

\vspace{0.3cm}

{\bf i.  Chiral quark model.} This is a nonrelativistic model which relates the
neutron EDM to the EDMs of the valence quarks with the help of the SU(6)
coefficients:
\begin{equation}
d_n={4\over 3} d_d -{1\over 3} d_u \;.
\end{equation} 
The quark EDMs can be $estimated$  via Naive Dimensional Analysis \cite{nda} as
\begin{equation}
d_q=\eta^E d_q^E + \eta^C {e\over 4\pi} d_q^C + \eta^G {e \Lambda \over 4\pi} d^G\;,
\end{equation}
where the QCD correction factors are given by $\eta^E=1.53$, $\eta^C \simeq \eta^G
\simeq 3.4$, and $\Lambda \simeq 1.19\;GeV$ is the chiral symmetry breaking scale.
We use the numerical values for these coefficients as given in \cite{nath}.
The parameters $\eta^{C,G}$ involve  considerable uncertainties steming from
the fact that the strong coupling constant at low energies is unknown.
Another weak side of the model is that it neglects the sea quark contributions 
which play an important role in the nucleon spin structure. 

The supersymmetric contributions to the dipole moments of the individual quarks
result from the 1-loop gluino, chargino, neutralino exchange diagrams
\begin{equation}
d_q^{E,C}=d_q^{\tilde g \;(E,C)}+d_q^{\chi^+ \; (E,C)} + d_q^{\chi^0 \; (E,C)}\;,
\end{equation}
and from the 2-loop gluino-quark-squark diagrams which generate $d^G$. \\

\vspace{0.2cm}

{\bf ii.  Parton quark model.}  This  model is based on the isospin 
symmetry and
known contributions of different quarks to the spin of the proton \cite{ellis}. 
The quantities $\Delta_q$ defined as $\langle n \vert {1\over 2}\bar q \gamma_{\mu}
\gamma_5 q \vert n \rangle = \Delta_q\; S_{\mu} $, where $S_{\mu}$ is the neutron
spin, are related by the isospin symmetry to the quantities $(\Delta_q)_p$ which
are measured in the deep inelastic scattering (and other) experiments, i.e.
$\Delta_u = (\Delta_d)_p$, $\Delta_d = (\Delta_u)_p$, and $\Delta_s = (\Delta_s)_p$.
To be exact, the neutron EDM depends on the (yet unknown) tensor charges rather than these axial charges.
The main  $assumption$ of the model is that the quark contributions to the NEDM are weighted by
the same factors $\Delta_i$, i.e.  \cite{ellis}
\begin{equation}
d_n=\eta^E (\Delta_d d_d^E + \Delta_u d_u^E +\Delta_s d_s^E)\;.
\end{equation}
In our numerical analysis we use the following values for these quantities
$\Delta_d=0.746$, $\Delta_u=-0.508$, and $\Delta_s=-0.226$ as they appear 
in the analysis of Ref.\cite{bartl}. 
As before, we have
\begin{equation}
d_q^{E}=d_q^{\tilde g \;(E)}+d_q^{\chi^+ \; (E)} + d_q^{\chi^0 \; (E)}\;.
\end{equation}
The major difference from the chiral quark model is a large strange quark
contribution (which is likely to be an overestimate \cite{pospelov}).
In particular, due to the large strange and charm quark masses,
the strange quark contribution dominates in most regions of the parameter space. 
This leads to considerable numerical differences between the predictions of the two
models.

The current experimental limit on the neutron EDM is \cite{bound}
\begin{equation}
d_n < 6.3 \times 10^{-26} {\rm e \; cm}\;.
\end{equation}

\subsection{Mercury EDM}

The EDM of the mercury atom results mostly from T-odd nuclear forces in the mercury 
nucleus \cite{Fischler:1992ha}, which induce the effective interaction  of the type 
$({\rm {\bf I}} \cdot \nabla ) \delta ( {\rm {\bf r}})$
between the electron and the nucleus of spin {\bf I} \cite{pospelov}.
In turn, the T-odd nuclear forces arise  due to the effective 4-fermion interaction
$\bar p p \bar n i \gamma_5 n$. It has been argued \cite{pospelov} that the mercury
EDM is primarily sensitive to the chromoelectric dipole moments of the quarks
and the limit \cite{mercury}
\begin{equation}
d_{Hg} < 2.1 \times 10^{-28} {\rm e \; cm}
\end{equation}
can be translated into
\begin{equation}
\vert d_d^C - d_u^C -0.012 d_s^C \vert /g_s < 7 \times 10^{-27} {\rm cm}\;,
\label{cedmlimits}
\end{equation}
where $g_s$ is the strong coupling constant.
As in the parton neutron model, there is a considerable strange quark contribution.
The relative coefficients of the  quark contributions in (\ref{cedmlimits}) are 
known better than those for the neutron, however the overall normalization is still
not free of uncertainties \cite{khrip}. 

\section{The EDMs in SUSY models}

We will study supersymmetric models with the following  high energy scale soft breaking 
potential
\begin{eqnarray}
 V_{SB} &=& m_{0\alpha}^2 \phi^*_{\alpha} \phi_{\alpha} -( B\mu H_1 H_2 +h.c.)
+ ( A_l Y_{ij}^l\; H_1 \tilde l_{Li} \tilde e_{Rj}^* + 
A_d Y_{ij}^d \; H_1 \tilde q_{Li} \tilde d_{Rj}^* \nonumber\\
&-& A_u Y_{ij}^u \;  H_2 \tilde q_{Li} \tilde u_{Rj}^* +h.c.) 
+ {1\over 2} ( m_3 \overline {\tilde {g}} \tilde g + m_2 \overline {\tilde {W^a}} \tilde W^a
+ m_1 \overline {\tilde {B}} \tilde B)\;,
\end{eqnarray}
where $\phi_{\alpha}$ denotes all the scalars of the theory. We generally allow for 
$A_l \not = A_u \not = A_d$ as well as  nonuniversal
gaugino and scalar masses, which is important for the analysis of D-brane models.
The $\mu$, $B$, $A_{\alpha}$, and $m_i$  parameters can be complex, however two of their phases
can be eliminated by the $U(1)_R$ and $U(1)_{PQ}$ transformations under which  these
parameters behave as spurions. 
The Peccei-Quinn transformation acts on the Higgs doublets and the ``right-handed'' superfields
in such a way that all the interactions but those which mix the two doublets are invariant.
The Peccei-Quinn charges are $Q_{PQ}(\mu)= Q_{PQ}(B \mu), \;Q_{PQ}(A)=Q_{PQ}(m_i)=0$.
The $U(1)_R$ transforms the Grassmann variable $\theta \rightarrow \theta e^{i\alpha}$ and the
fields in such a way that the integral of the superpotential over the Grassmann variables
is invariant, i.e. the $U(1)_R$ charge of the superpotential is 2. As a result,
$Q_{R}(B\mu)= Q_{R}( \mu)-2, \;Q_{R}(A)=Q_{R}(m_i)=-2$.
 The six physical
CP-phases of the theory are invariant under both $U(1)_R$ and $U(1)_{PQ}$, and can be
chosen as
\begin{equation}
Arg(A_d^* m_i) \;,\; Arg((B\mu)^*\mu A_{\alpha})\;,
\label{phases}
\end{equation}
where $i=1,2,3$ and $\alpha=d,u,l$. All other CP-phases can be expressed as their linear
combinations. If the A-terms are universal, there are four physical phases 
$Arg(A^* m_i) \;,\; Arg((B\mu)^*\mu A)$.

It is customary to choose the phase convention in which the Higgs potential parameter
$B\mu$ is real. In this case, the physical phases become $Arg(A_d^* m_i) $ and
$Arg(\mu A_{\alpha})$. If universality is assumed, the number of physical phases
reduces to two. In what follows we will set $m_2$ to be real by a $U(1)_R$ rotation.

Nonuniversality will play a crucial role in the D-brane and flavour models' analysis, but 
otherwise does not lead to  different conclusions for the models we study. Thus we will
assume  universal A-terms and gaugino masses unless otherwise specified.

In what follows we use $\tan\beta$, $m_0$, $A$, $m_i$ as input parameters and obtain
low energy quantities via the MSSM renormalization group equations (RGE).
We also  assume radiative electroweak symmetry breaking, i.e. that the magnitude
of the $\mu$ parameter is given (at tree level) by 
\begin{equation}
\vert \mu \vert^2 = {m^2_{H_1}-m^2_{H_2} \tan^2\beta \over \tan^2\beta -1} - {1\over 2} m_Z^2    \;.
\end{equation}
The phase of  $\mu$  is an input parameter and is RG-invariant. Our numerical results are sensitive
to the quark   masses which we fix at the $M_Z$ scale to be: $m_{u_i}=(0.005,1.40,165)$ GeV and
$m_{d_i}=(0.010,0.194,3.54)$ GeV. 
The light  quark masses are poorly determined  (in fact $m_u=0$
is not excluded) which results  in the uncertainties of
the EDM normalization; for definiteness, we have chosen the light quark
masses as they appear in \cite{nath}.
The GUT scale is assumed to be $2\times 10^{16}$ GeV.

It is well known that ${\cal{O}}(1)$ supersymmetric CP phases generally lead to unacceptably large
electric dipole moments which  constitutes the SUSY CP problem. In this paper we consider  different 
mechanisms for suppressing EDMs and analyze them in detail.

\subsection{Leading SUSY contributions to the EDMs}

\begin{figure}[ht]
\epsfig{figure=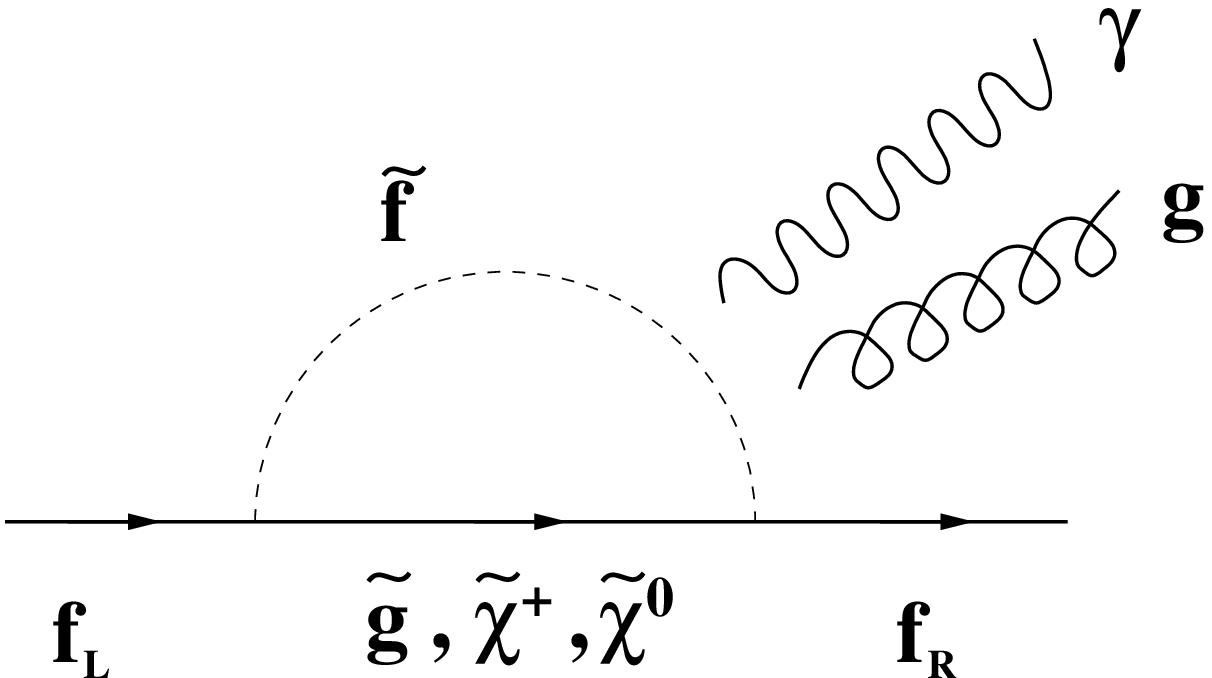,height=5cm,width=7cm,angle=0}
\hspace{2.cm}\epsfig{figure=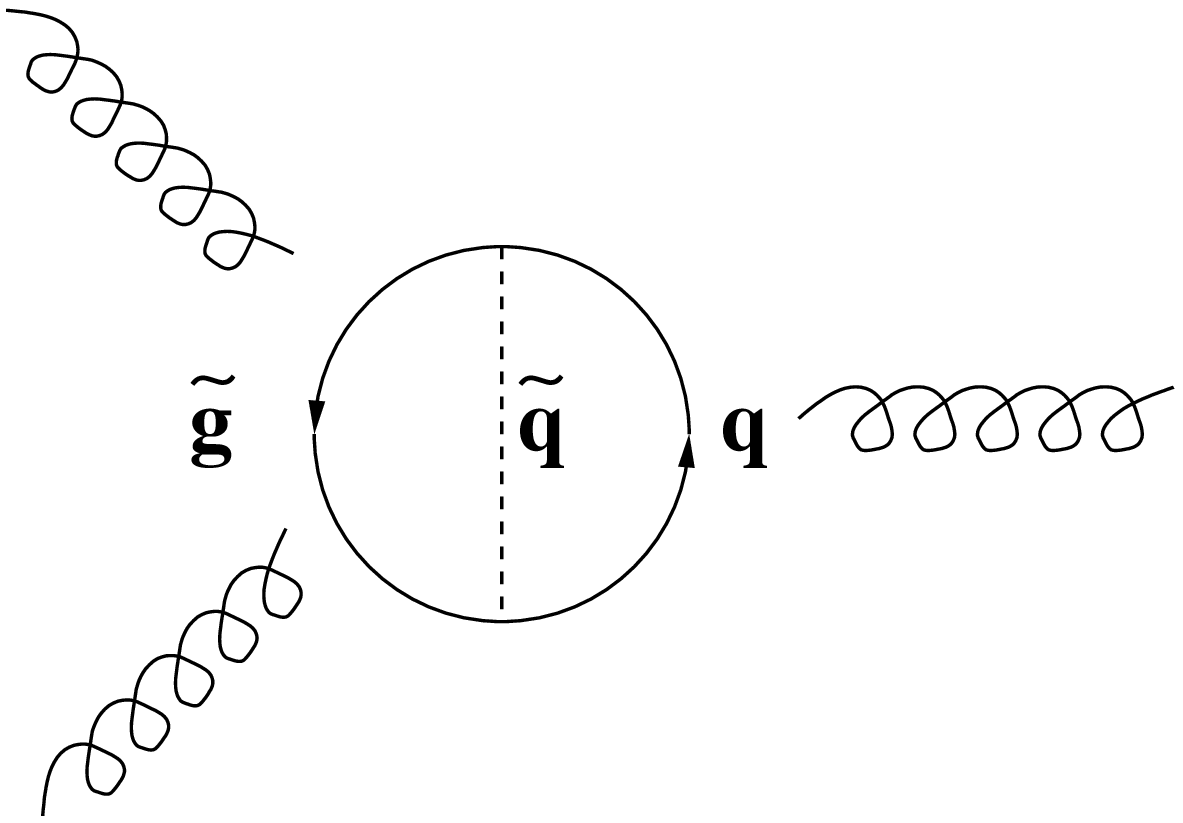,height=5cm,width=7cm,angle=0}
\medskip
\caption{Leading SUSY contributions to the EDMs. The photon and gluon lines are to be attached to
the loop in all possible ways.}
\label{diagram1}
\end{figure}

In this subsection we list formulae for individual supersymmetric contributions to the EDMs 
due to the Feynman diagrams in Fig.\ref{diagram1}.
In our presentation we follow the work of Ibrahim and Nath \cite{nath}.

Neglecting the flavour mixing, the electromagnetic contributions to the fermion EDMs are given by \cite{nath}:
\begin{eqnarray}
&& d_q^{\tilde g \; (E)}/e =  \frac{-2 \alpha_{s}}{3 \pi} 
\sum_{k=1}^2  
{\rm Im}(\Gamma_{q}^{1k})  \frac{ m_{\tilde{g}}}{M_{\tilde{q}_k}^2} Q_{\tilde{q}}\;
 {\rm B}\biggl( \frac{m_{\tilde{g}}^2}{M_{\tilde{q}_k}^2}\biggr)  \;,\nonumber\\
&& d_u^{\chi^+ \; (E)}/e = \frac{-\alpha_{em}}{4\pi\sin^2\theta_W}\sum_{k=1}^{2}\sum_{i=1}^{2}
      {\rm Im}(\Gamma_{uik})
               \frac{m_{\chi^+_i}}{M_{\tilde{d}_k}^2} \biggl[ Q_{\tilde{d}}\;
                {\rm B} \biggl( \frac{m_{\chi^+_i}^2}{M_{\tilde{d}_k}^2} \biggr)+
	(Q_u-Q_{\tilde{d}})\; {\rm A}\biggl( \frac{m_{\chi^+_i}^2}{M_{\tilde{d}_k}^2}\biggr) \biggr] \;,\nonumber\\
&& d_d^{\chi^+ \; (E)}/e = \frac{-\alpha_{em}}{4\pi\sin^2\theta_W}\sum_{k=1}^{2}\sum_{i=1}^{2}
      {\rm Im}(\Gamma_{dik})
               \frac{m_{\chi^+_i}}{M_{\tilde{u}_k}^2} \biggl[ Q_{\tilde{u}}\;
                {\rm B} \biggl( \frac{m_{\chi^+_i}^2}{M_{\tilde{u}_k}^2} \biggr)+
	(Q_d-Q_{\tilde{u}})\; {\rm A}\biggl( \frac{m_{\chi^+_i}^2}{M_{\tilde{u}_k}^2}\biggr) \biggr] \;,\nonumber\\
&& d_e^{\chi^+}/e=\frac{\alpha_{em}}{4\pi\sin^2\theta_W} 
      \sum_{i=1}^{2} {m_{\chi^+_i} \over  {m_{\tilde{\nu}}^2}} {\rm Im}
     (\Gamma_{ei})\;   
	{\rm A}\biggl( \frac{m_{\chi^+_i}^2}{m_{\tilde{\nu}}^2} \biggr)\;,\nonumber\\ 
&& d_f^{\chi^0 \; (E)}/e =\frac{\alpha_{em}}{4\pi\sin^2\theta_W}\sum_{k=1}^{2}\sum_{i=1}^{4}
{\rm Im}(\eta_{fik})
               \frac{m_{\chi^0_i}}{M_{\tilde{f}_k}^2} Q_{\tilde{f}}\;
{\rm B}\biggl( \frac{m_{\chi^0_i}^2}{M_{\tilde{f}_k}^2}\biggr) \;.
\end{eqnarray}
Here
\begin{equation}
\Gamma_{q}^{1k}=e^{-i\phi_3}D_{q2k}D_{q1k}^* \;,
\end{equation}
with $\phi_3$ being the gluino phase and $D_q$ defined by $D_q^{\dagger} M_{\tilde q}^2 D_q={\rm diag}( M_{\tilde q_1}^2, M_{\tilde q_2}^2)$. The sfermion mass matrix  $M_{\tilde f}^2$ is given by
\begin{eqnarray}
&&M_{\tilde{f}}^2=\left(\matrix{{M_L}^2+m{_f}^2+M_{z}^2(\frac{1}{2}-Q_f
\sin^2\theta_W)\cos2\beta & m_f(A_{f}^{*}-\mu R_f) \cr
m_f(A_{f} -\mu^{*} R_f) & M_{R
}^2+m{_f}^2+M_{z}^2 Q_f \sin^2\theta_W \;\cos2\beta}\right)\;,\nonumber
\end{eqnarray}
where $R_f=\cot\beta$ $(\tan\beta)$ for $I_3=1/2$ $(-1/2)$.
The chargino vertex $\Gamma_{fik}$ is defined as
\begin{eqnarray}
&&\Gamma_{uik}=\kappa_u V_{i2}^* D_{d1k} (U_{i1}^* D_{d1k}^*-
		\kappa_d U_{i2}^* D_{d2k}^*) \;,\nonumber\\
&&\Gamma_{dik}=\kappa_d U_{i2}^* D_{u1k} (V_{i1}^* D_{u1k}^*-
                \kappa_u V_{i2}^* D_{u2k}^*)
\end{eqnarray}
and analogously for the electron; here $U$ and $V$ are the unitary matrices diagonalizing the chargino mass
matrix: $U^* M_{\chi^+} V^{-1}= {\rm diag} (m_{\chi^+_1},m_{\chi^+_2})$. The quantities $\kappa_f$ are
the Yukawa couplings
\begin{equation}
\kappa_u=\frac{m_u}{\sqrt{2} m_W \sin\beta}, 
 ~~\kappa_{d,e}=\frac{m_{d,e}}{\sqrt{2} m_W \cos\beta}.
\end{equation}
The neutralino vertex $\eta_{fik}$ is given by
\begin{eqnarray}
\eta_{fik} & = &{\biggl[-\sqrt{2} \{\tan\theta_W (Q_f-I_{3_f}) X_{1i}
  +I_{3_f} X_{2i}\}D_{f1k}^*-
     \kappa_{f} X_{bi} D_{f2k}^*\biggr]}\nonumber\\
 &\times& {\biggl[ \sqrt{2} \tan\theta_W Q_f X_{1i} D_{f2k}
     -\kappa_{f} X_{bi} D_{f1k}\biggr]}\;,
\end{eqnarray}
where $I_3$ is the third component of the isospin,
$b=3\;(4)$ for $I_3=-1/2\;(1/2)$, and $X$ is the unitary matrix diagonalizing the 
neutralino mass matrix:  $X^T M_{\chi^0} X= {\rm diag} (m_{\chi^0_1},m_{\chi^0_2},m_{\chi^0_3},m_{\chi^0_4})$. In our convention the mass matrix eigenvalues are  positive and ordered as
$m_{\chi^0_1} > m_{\chi^0_2}>...$ (this holds for all mass matrices in the paper).
The loop functions $A(r), B(r)$, and $C(r)$ are defined by
\begin{eqnarray}
&& A(r)=\frac{1}{2(1-r)^2}\biggl(3-r+\frac{2\ln r}{1-r}\biggr) \;, \nonumber\\
&& B(r)=\frac{1}{2(r-1)^2}\biggl(1+r+\frac{2r\ln r}{1-r}\biggr)\; ,\nonumber\\
&& C(r)=\frac{1}{6(r-1)^2}\biggl(10r-26+\frac{2r\ln r}{1-r}-\frac{18\ln r}{1-r}\biggr)\;.
\end{eqnarray}

The chromoelectric contributions to the quark EDMs are given by
\begin{eqnarray}
&& d_q^{\tilde g \; (C)}=\frac{g_s\alpha_s}{4\pi} \sum_{k=1}^{2}
     {\rm Im}(\Gamma_{q}^{1k}) \frac{m_{\tilde{g}}}{M_{\tilde{q}_k}^2}\;
      {\rm C}\biggl(\frac{m_{\tilde{g}}^2}{M_{\tilde{q}_k}^2}\biggr)\;,\nonumber\\
&& d_q^{\chi^+ \; (C)}=\frac{-g^2 g_s}{16\pi^2}\sum_{k=1}^{2}\sum_{i=1}^{2}
      {\rm Im}(\Gamma_{qik})
               \frac{m_{\chi^+_i}}{M_{\tilde{q}_k}^2}\;
                {\rm B}\biggl(\frac{m_{\chi^+_i}^2}{M_{\tilde{q}_k}^2}\biggr)\;,\nonumber\\
&& d_q^{\chi^0 \; (C)}=\frac{g_s g^2}{16\pi^2}\sum_{k=1}^{2}\sum_{i=1}^{4}
       {\rm Im}(\eta_{qik})
               \frac{m_{\chi^0_i}}{M_{\tilde{q}_k}^2}\;
                {\rm B}\biggl(\frac{m_{\chi^0_i}^2}{M_{\tilde{q}_k}^2}\biggr)\;.
\end{eqnarray}

Finally, the contribution to the Weinberg operator \cite{Weinberg:1989dx} from
the two-loop gluino-top-stop and gluino-bottom-sbottom diagrams reads
\begin{equation}
d^G=-3\alpha_s m_t \biggl(\frac{g_s}{4\pi}\biggr)^3\;
{\rm Im} (\Gamma_{t}^{12})\;\frac{z_1-z_2}{m_{\tilde{g}}^3}\;
{\rm H}(z_1,z_2,z_t) +\;(t\rightarrow b)\;,
\end{equation}
where $z_{i}=\biggl(\frac{M_{\tilde{t}_i}}{m_{\tilde{g}}}\biggr)^2,
z_t=\biggl(\frac{m_t}{m_{\tilde{g}}}\biggr)^2$. The two-loop function $H(z_1,z_2,z_t)$
is given by \cite{Dai:1990xh}
\begin{equation}
H(z_1,z_2,z_t)={1\over2}\int_0^1 dx \int_0^1 du \int_0^1 dy\; x(1-x)u{N_1 N_2\over D^4}\;,
\end{equation}
where
\begin{eqnarray}
&& N_1=u(1-x)+z_t x(1-x)(1-u)-2ux[z_1y+z_2(1-y)]\;,\nonumber\\
&& N_2=(1-x)^2 (1-u)^2 +u^2-{1\over 9}x^2(1-u)^2 \;,\nonumber\\
&& D=u(1-x) +z_t x(1-x)(1-u)+ux[z_1y+z_2(1-y)]\;.
\end{eqnarray}
The numerical behaviour of this function was studied in \cite{Dai:1990xh}. We emphasize
that the b-quark contribution is significant and often exceeds the top one.

Before we proceed to the discussion of the EDM suppression mechanisms, let us  consider the effect of
other potentially nonnegligible two-loop contributions.  

\subsection{Barr-Zee type EDM  contributions}

In view of  considerable recent interest in the subject we will consider the two-loop Barr-Zee
type contributions separately. We will follow the work of Chang, Keung, and Pilaftsis \cite{chang}.

\begin{figure}[ht]
\epsfig{figure=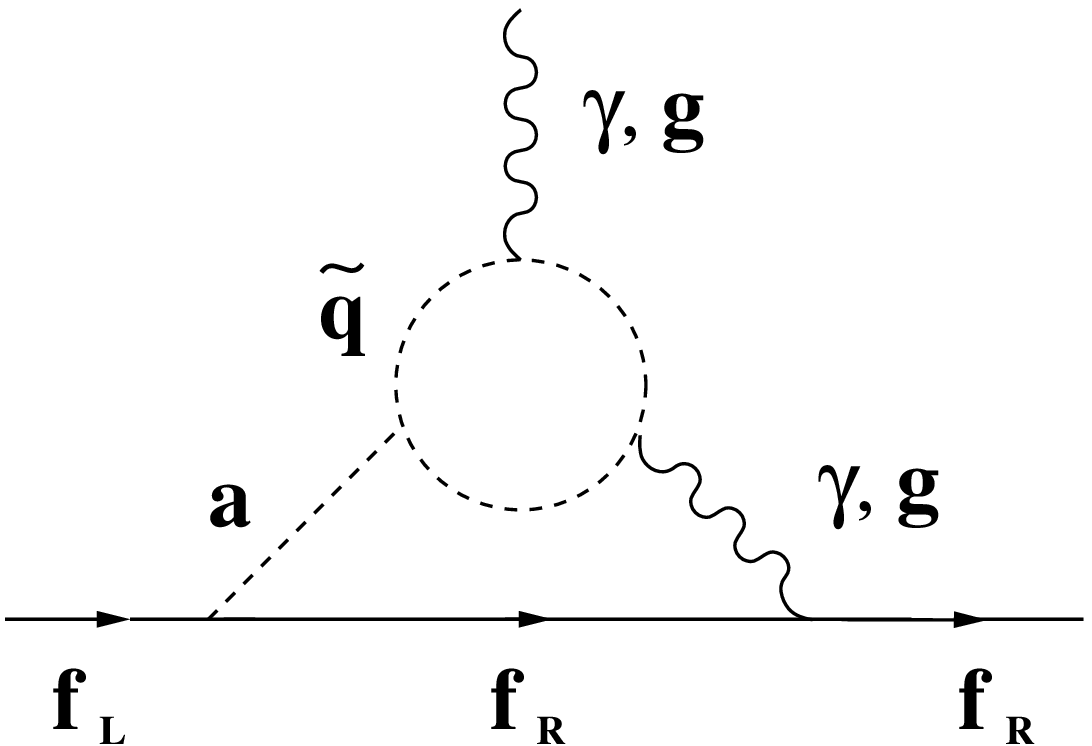,height=5cm,width=7cm,angle=0}
\hspace{2.cm}\epsfig{figure=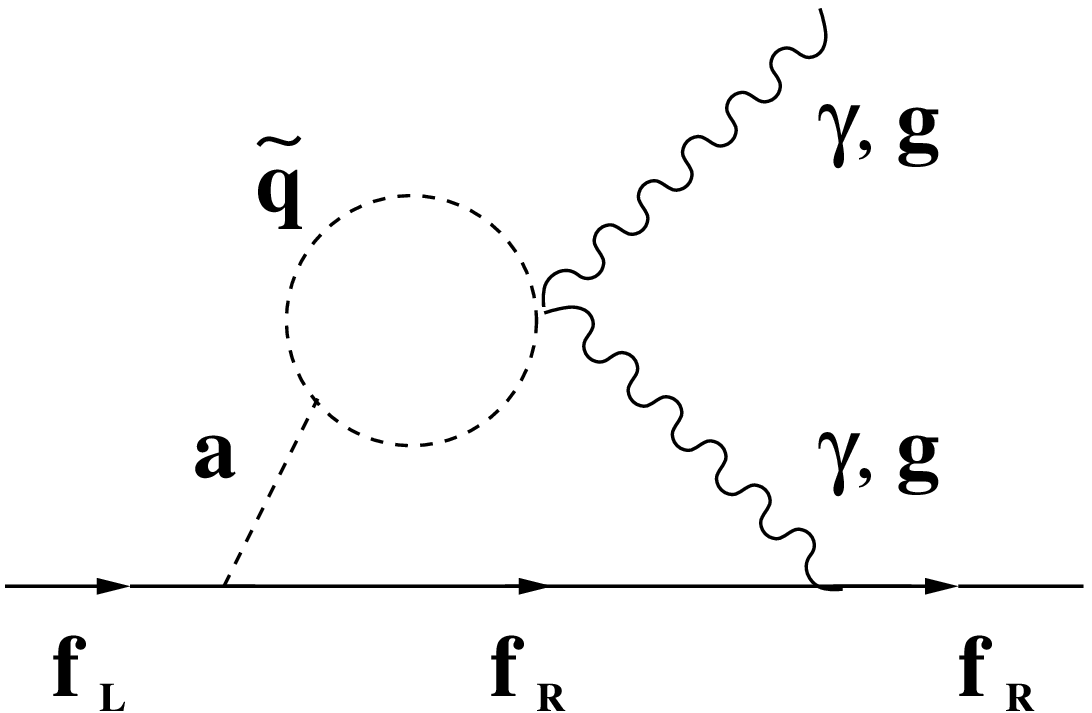,height=5cm,width=7cm,angle=0}
\medskip
\caption{Barr-Zee type contributions to the EDMs.}
\label{diagram2}
\end{figure}

In Ref.\cite{barr} Barr and Zee have presented two-loop Higgs-mediated EDM contributions
which can be competetive with the Weinberg three-gluon operator. A supersymmetric version of
the Barr-Zee graphs (Fig.\ref{diagram2}) was studied in \cite{chang}.
In what follows we will analyze only the leading contributions to the EDMs presented in \cite{chang}. 
The EDMs arise due to CP-violating couplings of the (s)fermions to the CP-odd Higgs boson $a_0$. 
The EDM and the CEDM of a light fermion $f$ are computed to be \cite{chang}
\begin{eqnarray}
&& d^E_f /e = Q_f \;{ 3 \alpha_{em} \over 32 \pi^3}\; {R_f m_f\over M_a^2}
\sum_{q=t,b} \xi_q Q_q^2 \biggl[ F \biggl( { M^2_{\tilde q_1} \over M_a^2} \biggr) - 
F \biggl( { M^2_{\tilde q_2} \over M_a^2} \biggr) \biggr] \;, \nonumber\\
&& d^C_f = { g_s \alpha_{s} \over 64 \pi^3}\; {R_f m_f\over M_a^2}
\sum_{q=t,b} \xi_q  \biggl[ F \biggl( { M^2_{\tilde q_1} \over M_a^2} \biggr) - 
F \biggl( { M^2_{\tilde q_2} \over M_a^2} \biggr) \biggr] \;,
\end{eqnarray}
where $R_f=\cot\beta$ $(\tan\beta)$ for $I_3=1/2$ $(-1/2)$ and  the two-loop function $F(z)$ is
\begin{equation}
F(z)=\int_0^1 dx {x(1-x) \over z-x(1-x)} \ln \biggl[  {x(1-x)\over z}   \biggr]\;.
\end{equation}
The CP-violating couplings $\xi_{t,b}$ are given by
\begin{eqnarray}
&& \xi_t =- { \sin 2\theta_{\tilde t} m_t {\rm Im}(\mu e^{i \delta_t}) \over 2 v^2 \sin^2\beta} \;, \nonumber \\
&& \xi_b = { \sin 2\theta_{\tilde b} m_b {\rm Im}(A_b e^{-i \delta_b}) \over 2 v^2 \sin\beta \cos\beta} \;,
\end{eqnarray}
with  $\theta_{\tilde t , \tilde b}$ being the standard stop and sbottom mixing angles;
$\delta_{q}={\rm Arg}(A_q - R_q\mu^*)$ and $v$=174 GeV. Note the difference from \cite{chang} in the
definitions of $\mu$ and $v$, also see the corresponding Erratum.

In our numerical analysis, besides assuming  the radiative electroweak symmetry breaking, we use the following 
tree level mass of  the CP-odd Higgs \cite{gunion}
\begin{equation}
M_a^2= m_{H_1}^2 + m_{H_2}^2 +2 \vert \mu \vert^2 \;,
\label{ma}
\end{equation}
which is a function of $\tan\beta$ and other GUT scale input parameters.
Strictly speaking, this formula is valid for a CP-conserving case, however
the EDMs are not very sensitive to the exact value of $M_a^2$ and an inclusion
of loop corrections and CP-phases does not alter our results.

\begin{figure}[ht]
\begin{center}
\begin{tabular}{c}
\epsfig{file=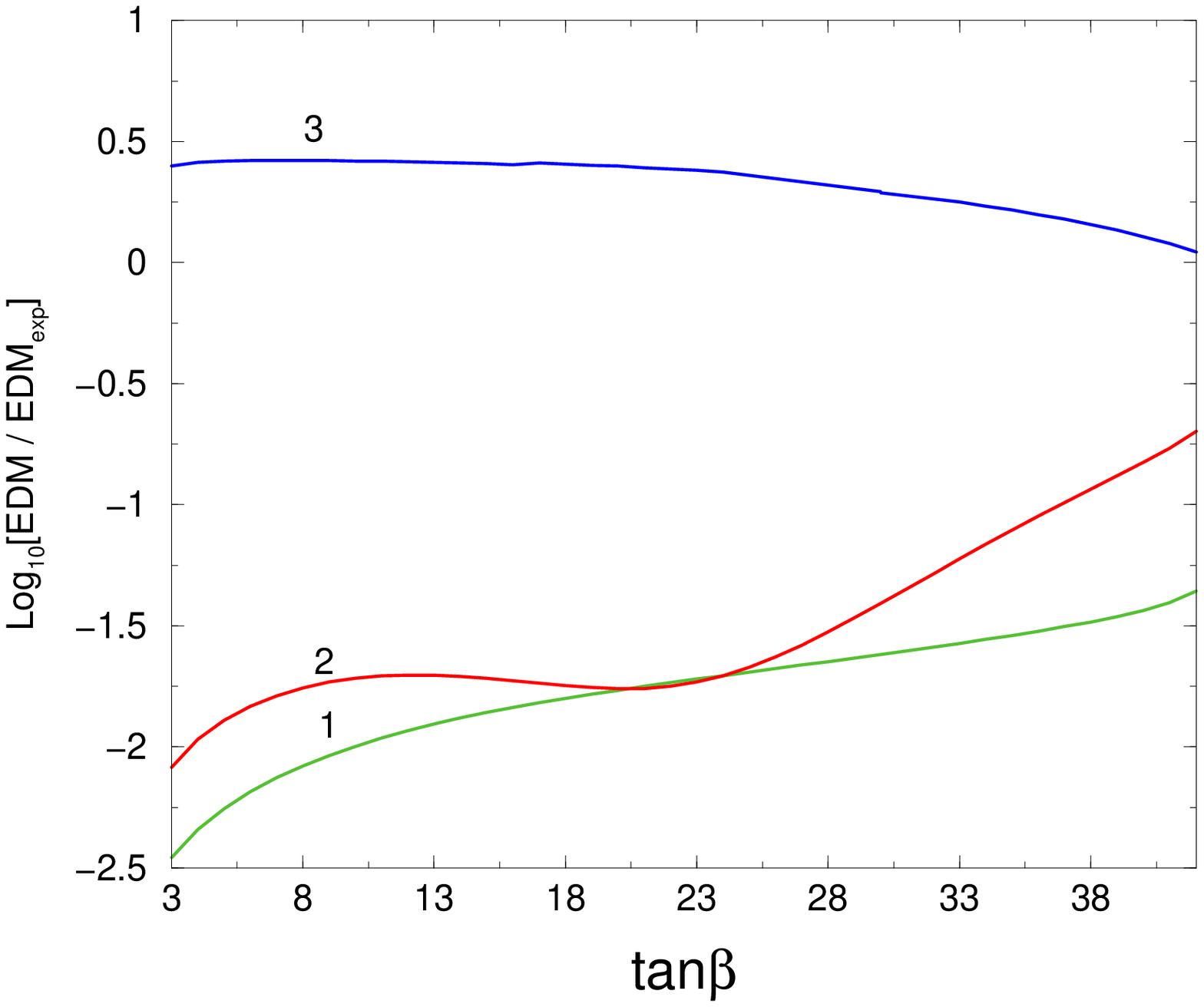, width=11cm, height=8.0cm}\\  
\end{tabular}
\end{center}
\caption{ Barr-Zee and Weinberg operator induced EDMs as a function
of $\tan\beta$. 1 -- electron Barr-Zee EDM,  2 -- neutron Barr-Zee EDM,
3 -- neutron EDM due to the Weinberg operator. Here $m_0=m_{1/2}=A=200$ GeV,
$\phi_{\mu}=0$, $\phi_A=\pi/2$. }
\label{barrzee1}
\end{figure}

In Fig.\ref{barrzee1} we present a typical Barr-Zee type EDM behaviour as a function of $\tan\beta$
for $\phi_A=\pi /2$, $\phi_{\mu}=0$. The other parameters are fixed to be
$m_0=m_{1/2}=A=200$ GeV.
The values of $\tan\beta$ beyond 42 are not displayed for this parameter set since 
the CP-odd scalar becomes massless in this region and the pattern of the EW symmetry breaking
becomes unacceptable.
We observe that generally these EDM contributions by themselves do not impose significant
constraints  on the GUT scale A-term phases even at large $\tan\beta$;
as can be seen from the plot, the Barr-Zee contributions typically are one-two orders of 
magnitude below the experimental limit. One of the reasons 
for this is that the third generation A-term phases reduce by an order of magnitude due to the 
RG evolution at large $\tan\beta$. Also the value of the $\mu$ parameter is typically below
500 GeV owing to the  radiative electroweak symmetry breaking. Other factors which distinguish
our results from those of  \cite{chang} are the imposition of Eq.(\ref{ma}) and utilization
of the chiral quark neutron model\footnote{ We observe  similar behaviour in the parton quark model.}.
For comparison, in Fig.\ref{barrzee1} we   provide the contribution 
of the Weinberg operator which also arises at the two-loop level. 

\begin{figure}[ht]
\begin{center}
\begin{tabular}{c}
\epsfig{file=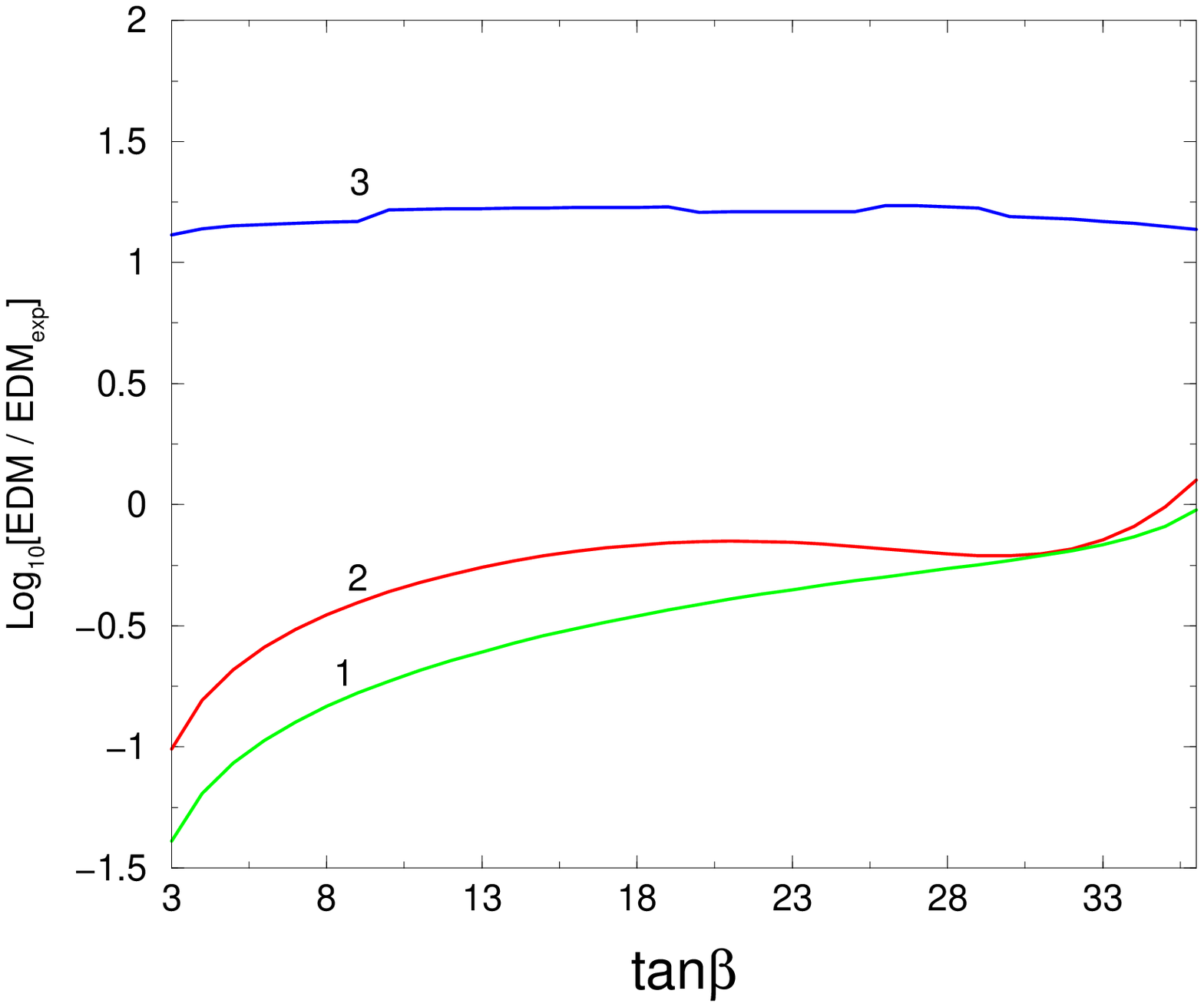, width=11cm, height=8.0cm}\\  
\end{tabular}
\end{center}
\caption{Barr-Zee and Weinberg operator induced EDMs as a function
of $\tan\beta$. 1 -- electron Barr-Zee EDM,  2 -- neutron Barr-Zee EDM,
3 -- neutron EDM due to the Weinberg operator. Here $m_0=500$ GeV,
$m_{1/2}=200$ GeV, $A=600$ GeV,
$\phi_{\mu}=0$, $\phi_A=\pi/2$.}
\label{barrzee2}
\end{figure}

The constraints become more restrictive for larger A-terms ($\sim 3 m_0$) and larger $m_0/m_3$
ratios. In Fig.\ref{barrzee2} we set $m_0=500$ GeV, $m_{1/2}=200$ GeV, and  $A=600$ GeV. 
In this case the A-term phase is not  ``diluted'' as much as before and for 
 some  parameters
the Barr-Zee EDMs can be close to  the experimental limit. A similar effect can be achieved in models
with non-universal gaugino masses by introducing
a ${\cal O}(1)$ gluino phase. With the parameter set  of Fig.\ref{barrzee2}  the CP-odd Higgs
becomes unacceptably light around $\tan\beta \simeq 36$. 

>From the point of view of low energy theory, the Barr-Zee type contributions can provide useful
constraints on the phases of the third generation A-terms \cite{chang},\cite{baek}.
One can imagine a situation in which the first two generation CP-violating effects are suppressed
(as in the decoupling scenario), then the EDMs would constrain the third generation phases. 
We find however that typically the Weinberg three-gluon operator is considerably more sensitive to such phases 
and provides more severe constraints  even at large $\tan\beta$\footnote{The Weinberg operator
contribution can be suppressed by increasing the gluino mass. However, in mSUGRA 
the Barr-Zee type contributions will also get a suppression factor  due to the RG ``dilution'' of the phases
of the third generation A-terms.}. For example, for the parameters of Figs.\ref{barrzee1} and \ref{barrzee2}
 the contribution
of the Weinberg operator exceeds that of the Barr-Zee type graphs by one-two orders of magnitude.

Finally, the Z- and W-mediated Barr-Zee type graphs have been analyzed in \cite{Chang:2000zw}
and found to be significantly  smaller than those considered above.
A number of other subleading  two loop contributions such as the gluino CEDM-induced quark EDMs, etc.
have been studied   in \cite{Pilaftsis:2000bq}. 
 
Although taken into account, this entire class of diagrams is numerically 
unimportant  in our analysis. 

\section{Suppression of the EDMs in SUSY models}

\subsection{Small CP-phases}

\begin{figure}[ht]
\begin{center}
\begin{tabular}{c}
\epsfig{file=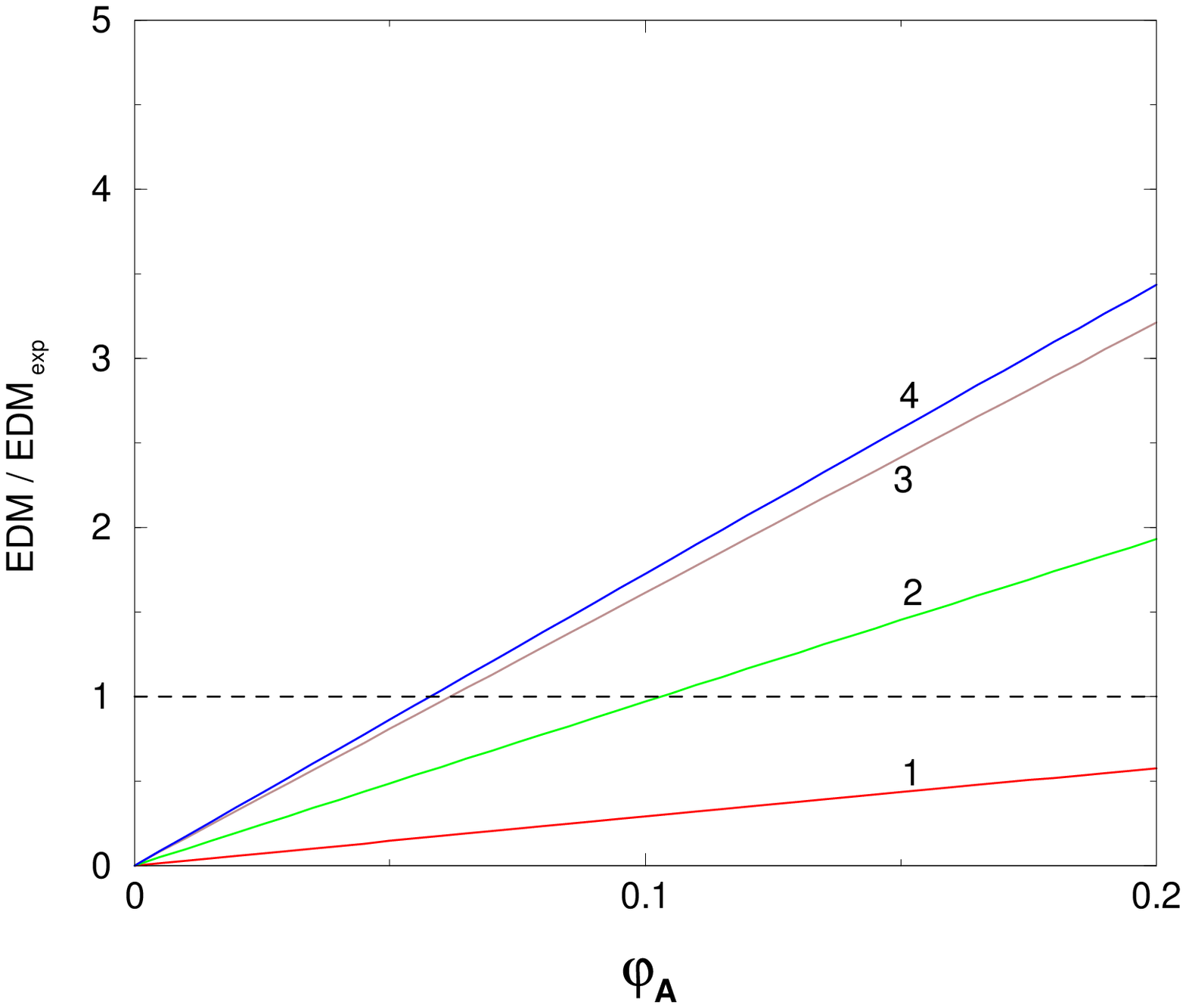, width=11cm, height=8.0cm}\\  
\end{tabular}
\end{center}
\caption{EDMs as a function of $\phi_A$.
 1 -- electron,  2 -- neutron (chiral model),
3 -- neutron (parton model), 4 -- mercury.
The experimental limit is given by the horizontal line. 
 Here $\tan\beta=3$, $m_0=m_{1/2}=A=200$ GeV.}
\label{aphase}
\end{figure}

\begin{figure}[ht]
\begin{center}
\begin{tabular}{c}
\epsfig{file=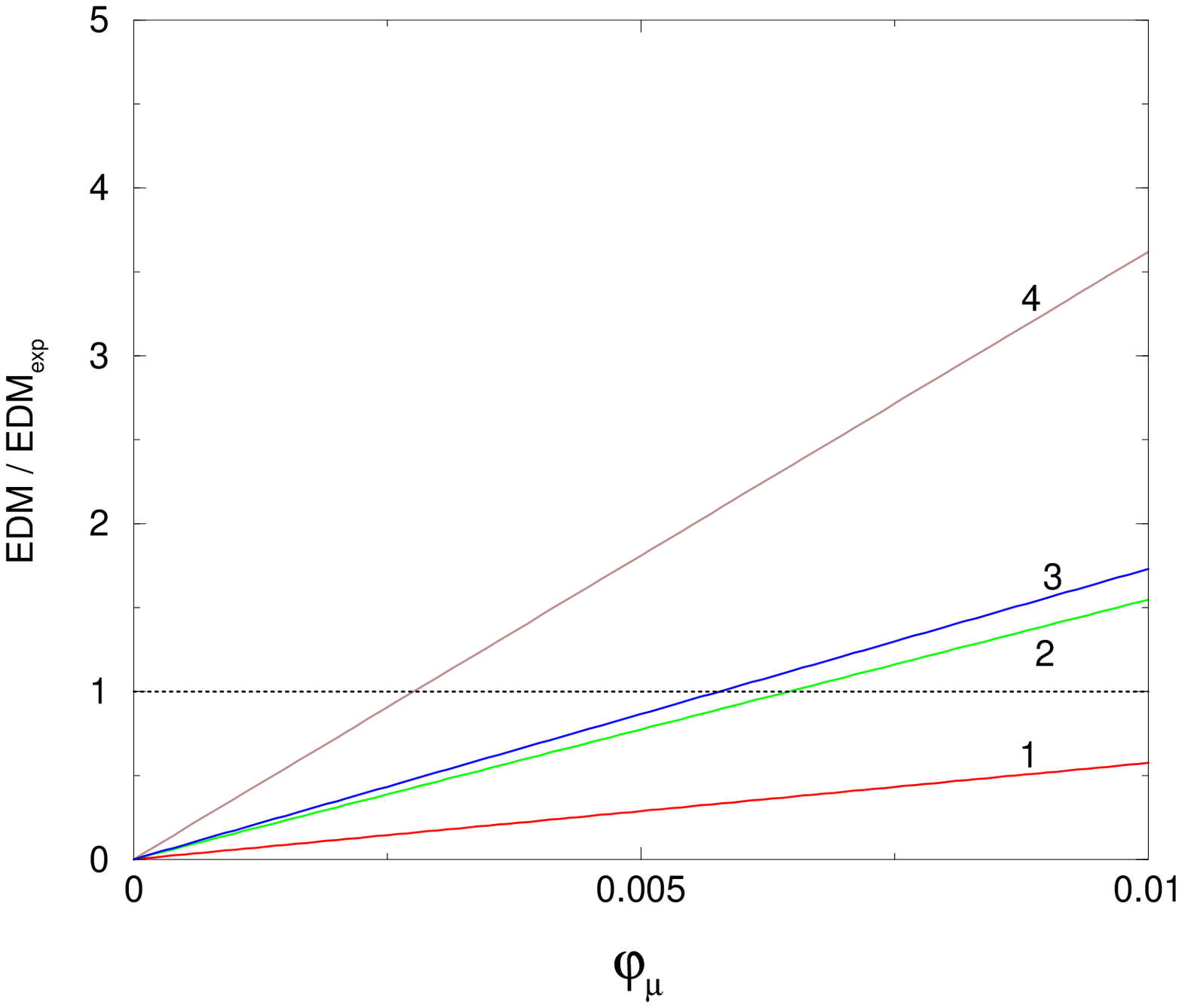, width=11cm, height=8.0cm}\\  
\end{tabular}
\end{center}
\caption{EDMs as a function of $\phi_{\mu}$.
 1 -- electron,  2 -- neutron (chiral model),
3 -- mercury, 4 -- neutron (parton model). 
The experimental limit is given by the horizontal line.
 Here $\tan\beta=3$, $m_0=m_{1/2}=A=200$ GeV.}
\label{muphase}
\end{figure}

\begin{figure}[ht]
\begin{center}
\begin{tabular}{c}
\epsfig{file=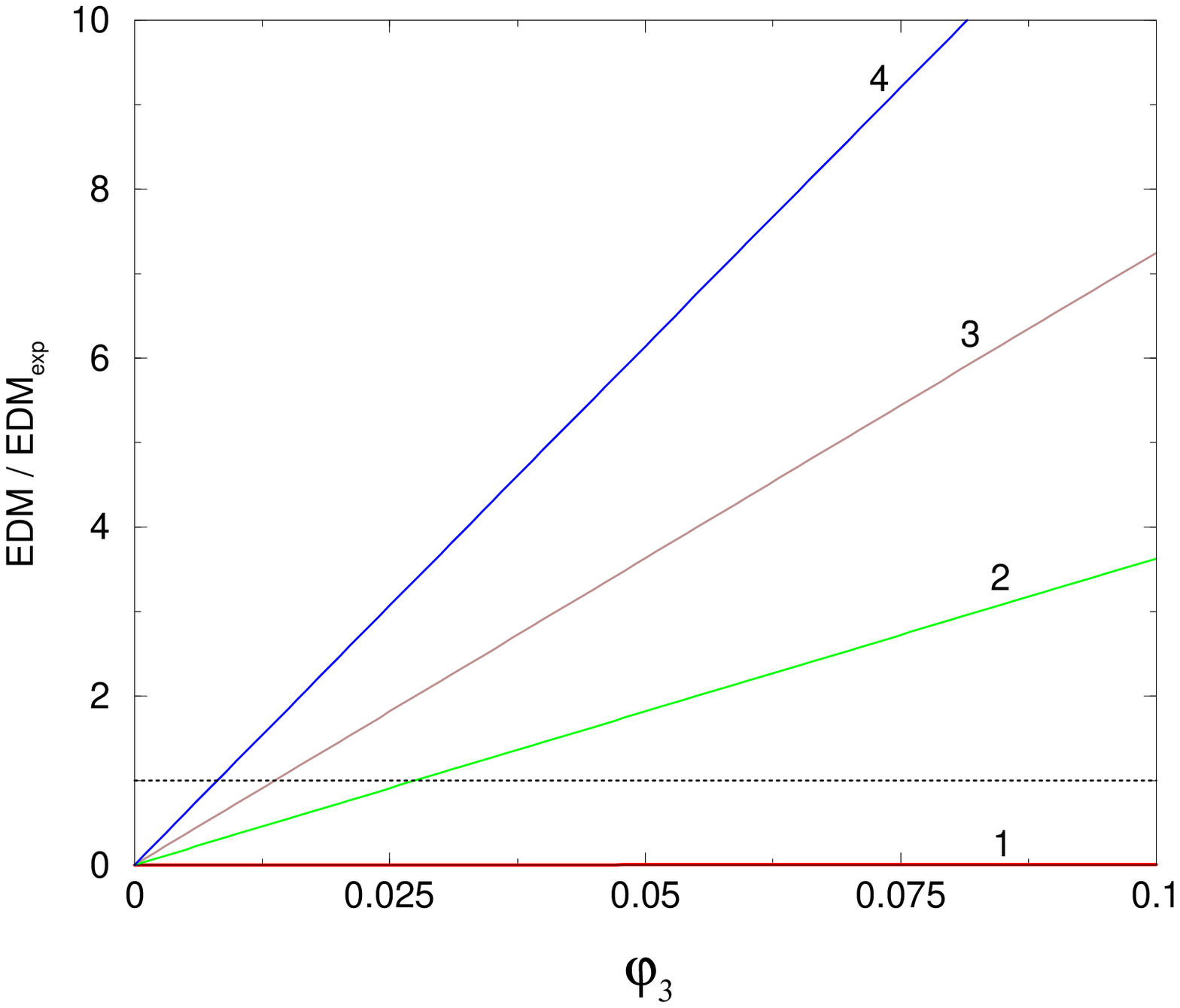, width=11cm, height=8.0cm}\\  
\end{tabular}
\end{center}
\caption{EDMs as a function of the gluino phase $\phi_{3}$.
 1 -- electron,  2 -- neutron (chiral model),
3 -- neutron (parton model), 4 -- mercury. 
The experimental limit is given by the horizontal line.
 Here $\tan\beta=3$, $m_0=m_{1/2}=A=200$ GeV.}
\label{gauginophase}
\end{figure}

For a light (below 1 TeV) supersymmetric spectrum,
 the SUSY CP phases have to be small
in order to satisfy the experimental EDM bounds (unless EDM cancellations occur). 
In Figs.\ref{aphase}-\ref{gauginophase} we illustrate the EDMs
behaviour as a function of the CP-phases in the mSUGRA-type models, where 
we have set $m_0=m_{1/2}=A=200$ GeV.  
At low $\tan\beta$, the EDM constraints impose the following bounds 
(at the GUT scale):
\begin{eqnarray}
&& \phi_A \leq 10^{-2}-10^{-1} \;, \nonumber\\
&& \phi_{\mu} \leq 10^{-3}-10^{-2} \;, \nonumber\\
&& \phi_{gaug.} \leq 10^{-2} \;.
\end{eqnarray}
For  $\tan\beta >3$, these bounds become even stricter (for 
 $\phi_{\mu}$ and $\phi_{gaug.}$ the bounds are roughly inversely
proportional to  $\tan\beta $). We note that $\phi_A$ is less constrained
than $\phi_{\mu}$ and $\phi_{gaug.}$. There are two reasons for that:
first, $\phi_A$ is reduced by the RG running from the GUT scale down to the
electroweak scale and, second, the phase of the $(\delta_{11}^d)_{LR}$
mass insertion which gives the dominant contribution to the EDMs is more 
sensitive to $\phi_{\mu}$  and $\phi_{gaug.}$ due to $\vert A \vert < \mu\tan\beta$.

SUSY models with small CP-phases can be motivated by the approximate CP 
symmetry \cite{nir}.
The well established experimental results exhibit small degree of CP violation.
Thus it is conceivable that all existing CP-phases are small, including the CKM
phase. The smallness of CP violation can be explained, for example, by a small
ratio of the scale at which the CP symmetry gets broken spontaneously and the
scale at which it is communicated to the observable sector.
 
Currently the CKM phase is consistent with zero and it could be supersymmetry that
is responsible for the observed values of $\varepsilon$ and $\varepsilon'$.
This does not require large supersymmetric phases. In fact 
in models with non-universal A-terms $\varepsilon$ and 
$\varepsilon'$ can even be saturated with ${\cal{O}}(10^{-2})$ phase of
the mass insertion $(\delta^d_{12})_{LR}$ \cite{eeprime}.
In this case, $\vert (\delta^d_{12})_{LR} \vert$ is required to be ${\cal{O}}(10^{-3})$
which naturally appears in models with matrix-factorizable  A-terms of the 
form $B\cdot Y_{\alpha} \cdot C$, where $B$ and $C$ are flavour matrices.  
The EDM bounds serve to  constrain the flavour structure of  $B$ and $C$. Another possibility
to produce the desired $(\delta^d_{12})_{LR}$ is to use asymmetric A-term textures
 in string-motivated  models (where the standard supergravity relation
$\hat A_{ij}=A_{ij} Y_{ij}$ is assumed).

Encouragingly, the phase of the $\mu$-term of order $10^{-2}$ may be sufficient to produce the
observed baryon asymmetry  \cite{carena},
which is in marginal agreement with the EDM bounds.
The hypothesis of the approximate CP symmetry is currently being tested in 
the B physics experiments where  the Standard Model predicts large CP-asymmetries.   
It is noteworthy that
the smallness of the CP-phases in this picture does not constitute fine-tuning 
according to the t'Hooft's criterion \cite{thooft} since setting them to zero would 
increase the symmetry of the theory.

We remark that small CP-phases may also arise due to the dynamics of the system.
For instance, in weakly coupled heterotic string models, small soft and CKM phases
arise when the T-moduli get VEV's close to the edge of the fundamental domain
which is often the case \cite{Bailin:1998iz}. 
This mechanism however relies on the assumption that the dilaton has a real VEV, so this
model as it stands does not solve the SUSY CP problem. Nevertheless, it may serve as a  
step toward a consistent string model with naturally small CP phases. 

Note that EDMs constrain only the  $physical$ phases (\ref{phases}).
One can imagine a situation when the individual phases are ${\cal{O}}(1)$ whereas 
the physical ones are small. This occurs for example in gauge mediated SUSY breaking
models,  see \cite{nir} and references therein.

\subsection{Heavy SUSY scalars}

This possibility is based on the decoupling of heavy supersymmetric particles.
Even if one allows ${\cal{O}}(1)$ CP violating phases, their effect will be
negligible if the SUSY spectrum is sufficiently heavy \cite{heavy}. Generally, SUSY fermions
are required to be lighter than the SUSY scalars by, for example, cosmological
considerations. So the decoupling scenario can be implemented with heavy
sfermions only.
Here the SUSY contributions to the EDMs are suppressed even
with maximal SUSY phases  because the squarks in the loop 
are very heavy and  the mixing angles are  small.

\begin{figure}[ht]
\begin{center}
\begin{tabular}{c}
\epsfig{file=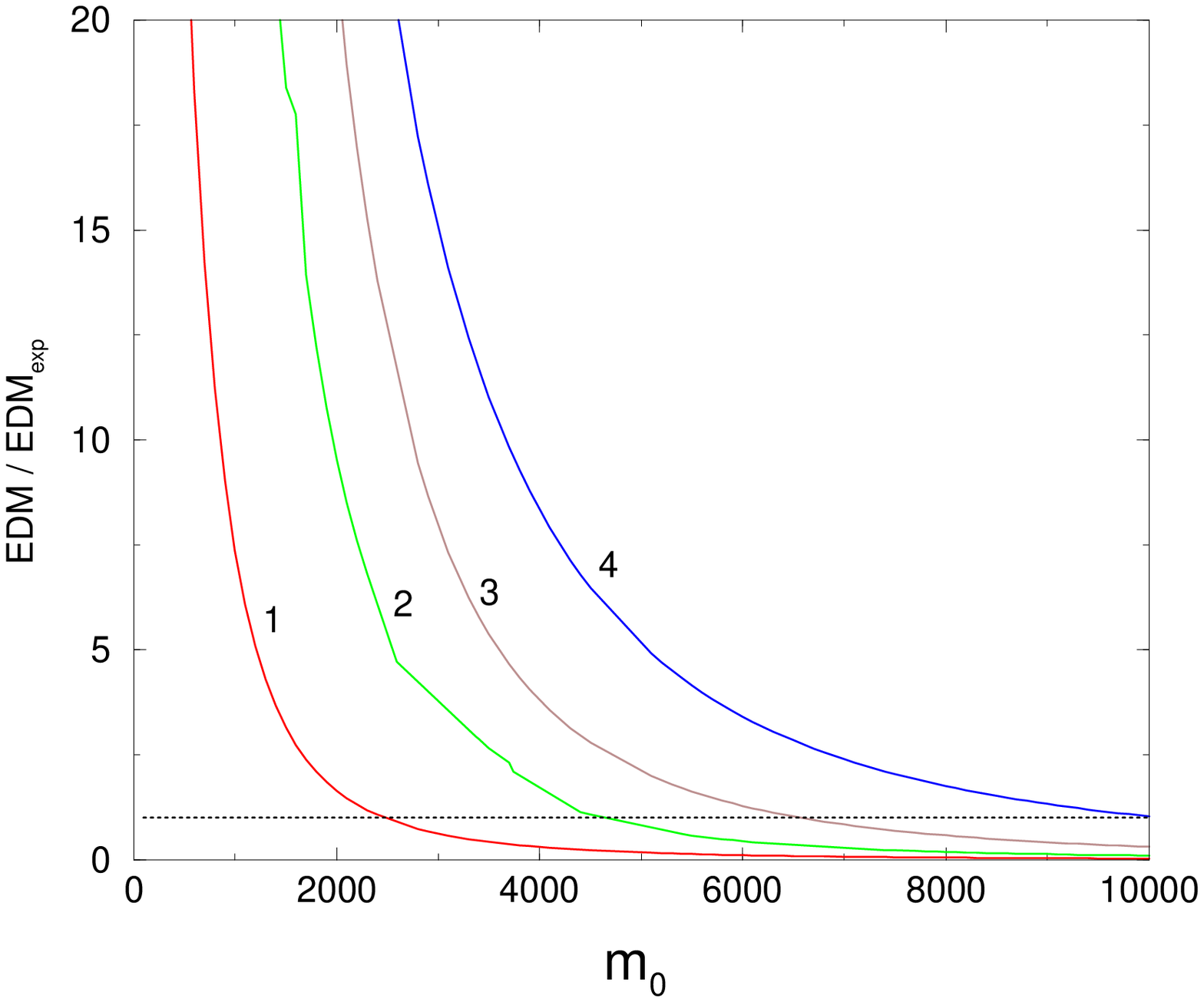, width=11cm, height=8.0cm}\\  
\end{tabular}
\end{center}
\caption{EDMs as a function of the universal mass parameter
$m_0$. 1 -- electron,  2 -- neutron (chiral model),
3 -- neutron (parton model), 4 -- mercury. 
The experimental limit is given by the horizontal line.
Here  $\tan\beta=3$, $m_{1/2}=A=200$ GeV, 
$\phi_{\mu}=\phi_A=\pi/2$.}
\label{heavy}
\end{figure}

In Fig.\ref{heavy} we display the EDMs as functions of the universal scalar mass parameter $m_0$
for the mSUGRA model with maximal CP-phases $\phi_{\mu}=\phi_A=\pi/2$ and
$m_{1/2}=A=200$ GeV. 
We observe that all EDM constraints except for that of the electron require $m_0$ to
be around 5 TeV or more. The mercury constraint is the strongest one and 
requires
\begin{equation}
(m_0)_{decoupl.} \simeq 10\;{\rm TeV}\;.
\end{equation} 
This leads to a serious fine-tuning problem. Recall that one of the primary motivations
for supersymmetry was a solution to the naturalness problem. Certainly this motivation
will be entirely lost if a SUSY model reintroduces the same problem in a different 
sector, i.e. for example a large hierarchy between the scalar mass and the electroweak 
scale. 

The degree of fine-tuning can be quantified as follows. The Z boson mass is determined
at tree level by
\begin{equation}
{1\over 2} m_Z^2= {m^2_{H_1}-m^2_{H_2} \tan^2\beta \over \tan^2\beta -1} -\mu^2\;.
\end{equation}
One can define the sensitivity coefficients \cite{finetuning},\cite{focus}
\begin{eqnarray}
&& c_i \equiv \biggl\vert {\partial \ln m_Z^2 \over \partial \ln a_i^2 }\biggr\vert \;,
\label{sens}
\end{eqnarray}
where $a_i$ are the high energy scale input parameters such as $m_{1/2}$, $m_0$, etc.
Note that $\mu$ is treated here as an independent input parameter.
A value of $c_i$ much greater than one would indicate a large degree of fine-tuning.
The Higgs mass parameters are quite sensitive to $m_0$, so for $m_0$=10 TeV we find
\begin{equation}
c_{m_0} \sim 5000
\end{equation}
for the parameters of Fig.\ref{heavy}. For the universal scalar mass of 5 (3) TeV the sensitivity
coefficient reduces to 1300 (500). This clearly indicates an unacceptable degree of
fine-tuning.

This problem can be mitigated in models with focus point supersymmetry, i.e. when
$m_{H_2}$ is insensitive to $m_0$ \cite{focus}. However, this mechanism works for $m_0$
no greater than 2-3 TeV which is not sufficient to suppress the EDMs.
Another interesting possibility is presented by models with a radiatively driven inverted
mass hierarchy, i.e. the models in which a large hierarchy between the Higgs and the first two
generations scalar masses is created radiatively \cite{Feng:1999iq}. 
However, a successful implementation of this idea is far from trivial \cite{baer}. 
One can also  break the scalar mass universality at the high energy scale \cite{dim}.
In this case,  either  a  mass hierarchy  appears already in the soft breaking terms
or certain relations among the soft parameters must be imposed (for a review see \cite{Baer:2000gf}).
These significant complications  disfavour the decoupling scenario as a way to solve
the SUSY CP problem, yet it remains a possibility. 

Note  that the decoupling scenario rules out supersymmetry as a possible explanation 
of the recently  observed 2.6 $\sigma$ deviation in the muon $g-2$ from the SM prediction
\cite{muon}.  This scenario may also lead to cosmological difficulties, in particular 
with the relic abundance of the LSPs since the LSP annihilation cross section falls
rapidly as $m_{sfermion}$ increases. 
Concerning the other phenomenological consequences, we remark that 
the SUSY contributions to the CP-observables involving the first two generations
 (such as $\varepsilon, \varepsilon'$) are negligible, although those involving the third generation
may be considerable. The corresponding CP-phases are constrained through the Weinberg operator 
contribution to the neutron EDM, which typically prohibits the maximal phase 
$\phi_{A_{t,b}}\Bigl\vert_{GUT}=  \pi/2$ while
still allowing for smaller ${\cal O}(1)$ phases (Fig.\ref{barrzee1}). 

\subsection{EDM cancellations}

The cancellation scenario is based on the fact that large cancellations among different
contributions to the EDMs are possible in certain regions of the parameter space 
\cite{nath},\cite{cancel},\cite{savoy},\cite{bartl} which allow for ${\cal{O}}(1)$ 
$flavour-independent$ CP phases.
Although not particularly well motivated, this possibility is interesting since, 
when supplemented with a  $real$ non-trivial flavour structure,
it allows for 
significant supersymmetric contributions to CP observables including those in the B system. 
In this case, the SM predictions can be significantly altered 
leading for instance to a nonclosure of the unitarity triangle (see M. Brhlik {\it et al.} in
\cite{eeprime}). Given the appropriate A-terms' flavour structures,
 the parameters $\varepsilon$ and $\varepsilon'$ can be of completely
supersymmetric nature \cite{eeprime}.
Large flavour-independent SUSY phases  may also be responsible for
electroweak baryogenesis \cite{baryo}.
Thus the cancellation  scenario presents a interesting  alternative to the decoupling 
and approximate CP solutions. 

For the case of the electron, the EDM cancellations occur between the chargino and the neutralino
contributions.
For the case of the neutron and mercury, there are cancellations between the gluino and the chargino
contributions as well as cancellations among contributions of different quarks to the total
EDM. A number of approximate formulae quantifying  the cancellations are presented in Refs.\cite{nath},\cite{cancel}.

In what follows we examine the cancellation scenario in the universal and nonuniversal cases,
with the latter being motivated by Type I string models.
We note that  the CP phases are  to be understood modulo $\pi$.

\subsubsection{EDM cancellations in mSUGRA-type models.}

These are  mSUGRA-type models allowing different phases for different gaugino masses while keeping
their magnitudes the same. As we will see, mSUGRA models with zero gaugino phases 
can realize the cancellation scenario. An introduction of the gaugino phases
makes the cancellations much harder to achieve (if possible at all) and leads to a much higher degree 
of fine-tuning.
The parameters allowing the EDM cancellations strongly depend on the neutron model.
For example, in the parton model, it is more difficult to achieve these cancellations
due to the large strange quark contribution. Therefore, one cannot restrict the parameter
space in a model-independent way and caution is needed when dealing with the parameters
allowed by the cancellations. 

\begin{figure}[ht]
\begin{center}
\begin{tabular}{c}
\epsfig{file=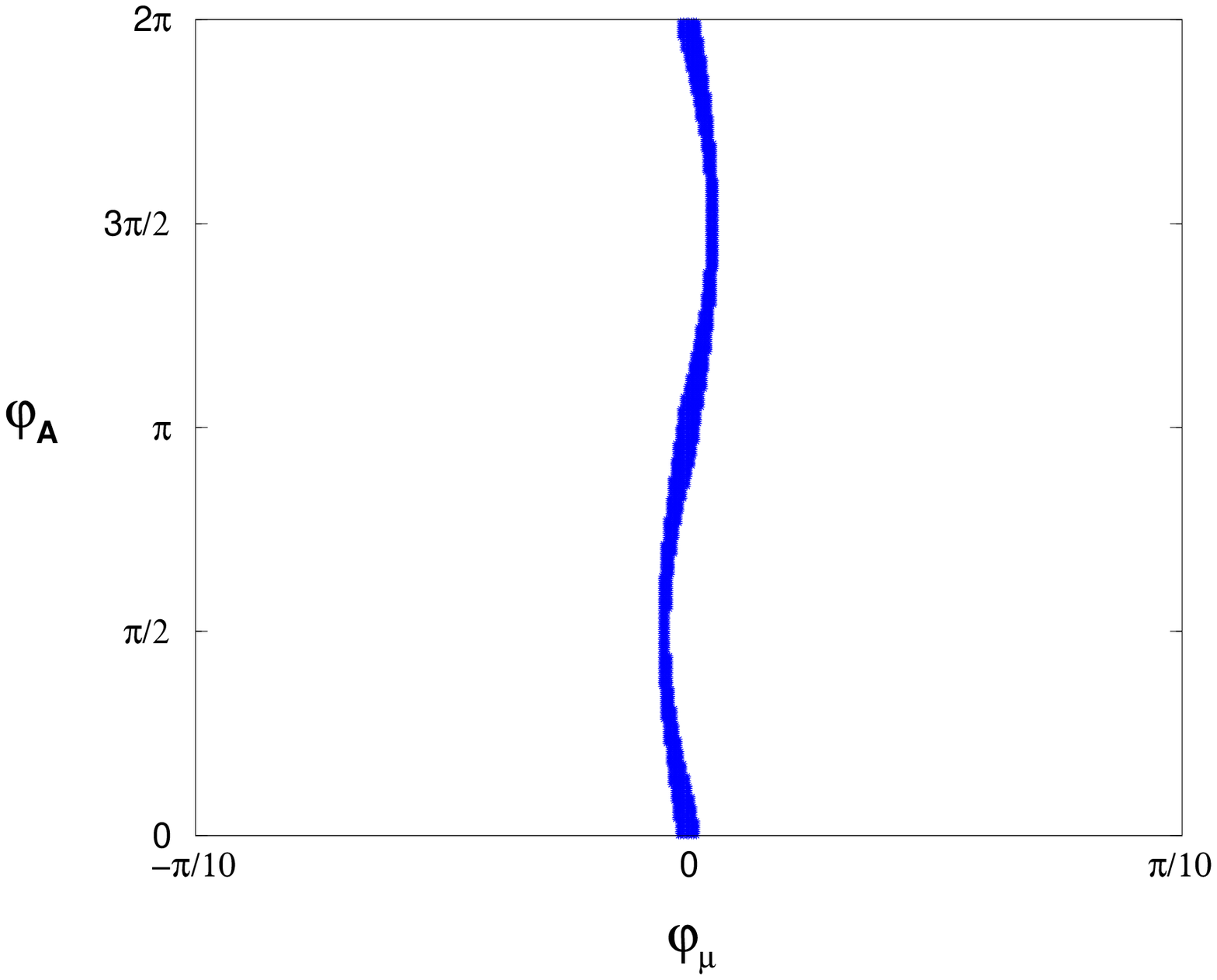, width=11cm, height=8.0cm}\\  
\end{tabular}
\end{center}
\caption{Phases allowed by simultaneous electron, neutron, and mercury EDM
cancellations in mSUGRA. The $chiral$ quark neutron model is assumed.
Here  $\tan\beta=3$, $m_0=m_{1/2}=200$ GeV, $A=40$ GeV. }
\label{sugra}
\end{figure}

\begin{figure}[ht]
\begin{center}
\begin{tabular}{c}
\epsfig{file=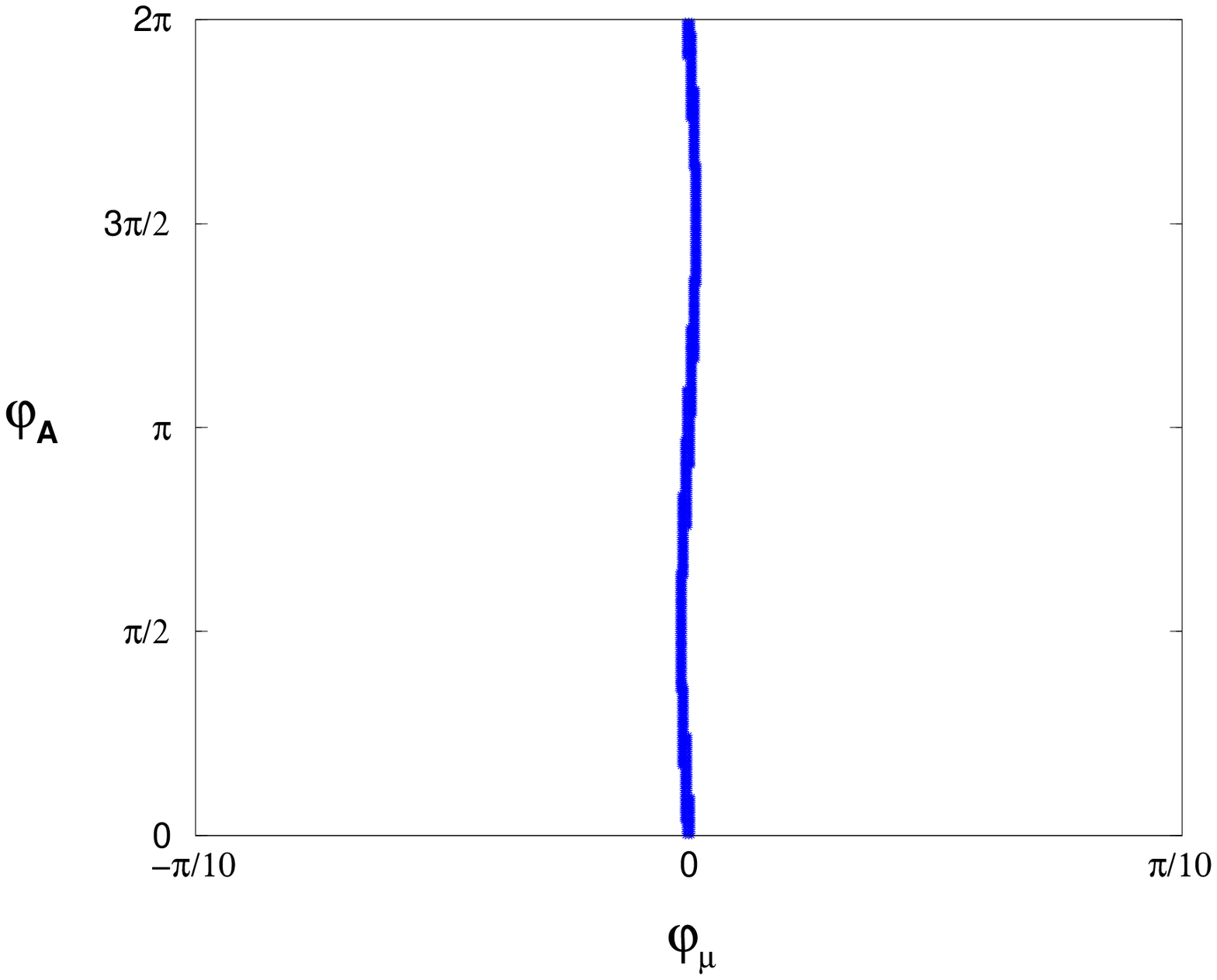, width=11cm, height=8.0cm}\\  
\end{tabular}
\end{center}
\caption{Phases allowed by simultaneous electron, neutron, and mercury EDM
cancellations in mSUGRA. The $parton$ quark neutron model is assumed.
Here  $\tan\beta=3$, $m_0=m_{1/2}=200$ GeV, $A=20$ GeV. }
\label{parton}
\end{figure}

\begin{figure}[ht]
\begin{center}
\begin{tabular}{c}
\epsfig{file=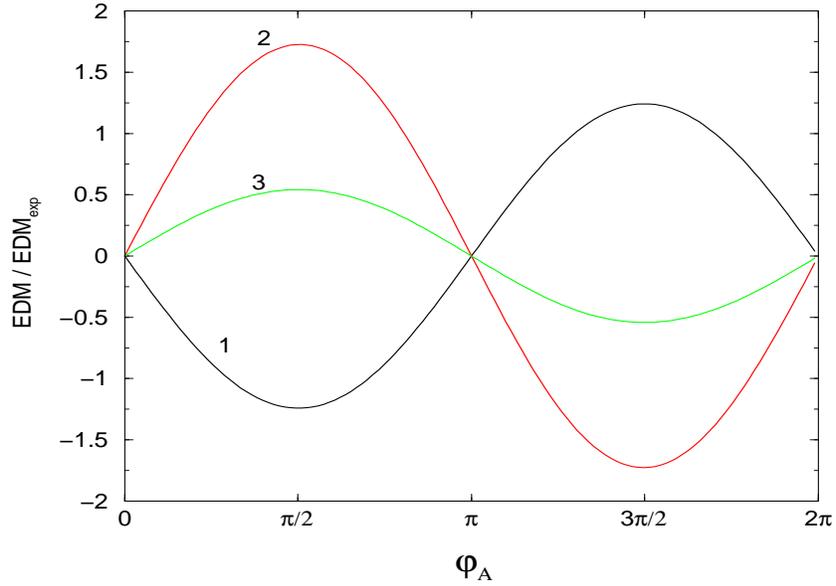, width=11cm, height=8.0cm}\\  
\end{tabular}
\end{center}
\caption{ Chargino - neutralino cancellations for the EEDM.
1 -- neutralino, 2 -- chargino, 3 -- total EDM.
The SUSY parameters are the same as for Fig.\ref{sugra},
which allow for simultaneous electron, neutron, and mercury EDM
cancellations. }
\label{electron}
\end{figure}

\begin{figure}[ht]
\begin{center}
\begin{tabular}{c}
\epsfig{file=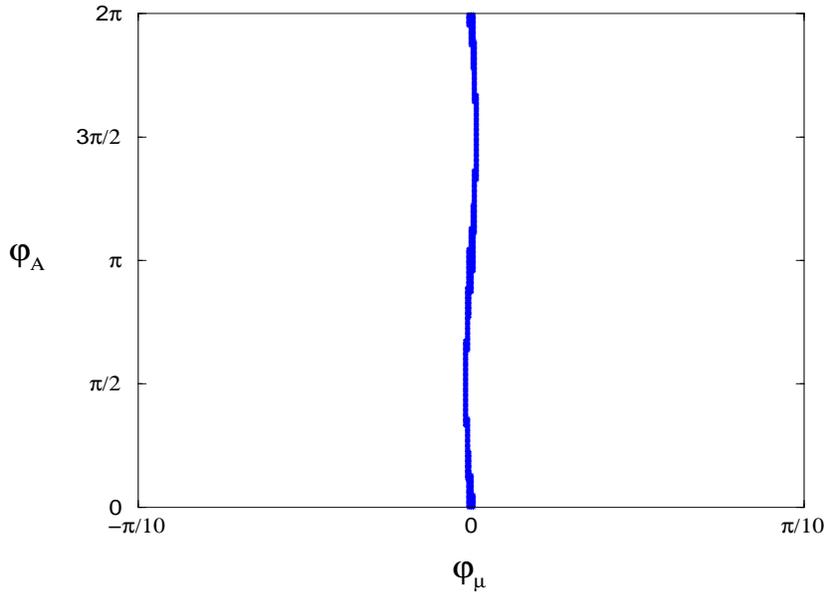, width=11cm, height=8.0cm}\\  
\end{tabular}
\end{center}
\caption{ Phases allowed by simultaneous electron, neutron (chiral model), 
and mercury EDM cancellations in mSUGRA at larger $\tan\beta$. 
Here  $\tan\beta=10$, $m_0=m_{1/2}=200$ GeV, $A=40$ GeV. }
\label{tanb10}
\end{figure}

In mSUGRA, the EDM  cancellations can occur simultaneously for the electron, neutron, 
and mercury
along a band in the $(\phi_A, \phi_{\mu})$ plane (Figs.\ref{sugra} and \ref{parton}).
However, in this case the mercury constraint requires the $\mu$ phase to be  ${\cal{O}}(10^{-2})$ 
and the magnitude of the A-terms to be suppressed ($\sim 0.1m_0$) 
which results in only a small effect of the A-terms on the phase of the corresponding
mass insertion.
 This is to be contrasted
with simultaneous EEDM/NEDM cancellations which allow for $\phi_{\mu}\sim{\cal{O}}(10^{-1})$
and $A \sim m_0$ \cite{bartl},\cite{cancel}. In that case the individual contributions
to the EDMs exceeded the experimental limit by an order of magnitude (or more). 
Owing to the addition of the mercury constraint, the EDM cancellations  become  much milder
as illustrated in  Fig.\ref{electron}.
For example,  without these cancellations  the EDMs   would exceed the experimental limit
only by a factor of a few. Obviously, the border between the cancellation and the small phases
scenarios becomes blurred. 
This is even more so at larger $\tan\beta$ (Fig.\ref{tanb10}), in which case the cancellation
band becomes  narrower and the limits on $\phi_{\mu}$ become tighter.
We note that there could exist some  points which allow
for $\phi_{\mu}\sim{\cal{O}}(10^{-1})$ \cite{barger}, however these are rare exceptions and 
such points do not form a band.

As noted in Ref.\cite{bartl}, in the case of zero gaugino phases, the band in the $(\phi_A, \phi_{\mu})$ 
plane where the cancellations occur can approximately be described by a relation
\begin{equation}
\phi_{\mu} \simeq - a \sin \phi_A\;,
\label{band}
\end{equation}
where $a$ is a constant depending on the parameters of the model which represents the maximal 
allowed phase $\phi_{\mu}$. For example, for
the chiral quark neutron model and 
$\tan\beta=3$, $m_0=m_{1/2}=200$ GeV, and $\vert A \vert=40$ GeV (parameters of
Fig.\ref{sugra}), $a=0.017$. Of course, the value of $a$ depends on the input parameters such as the
quark masses, the GUT scale, the SUSY scale, etc. and also involves numerical uncertainties, so
caution is needed when treating this number. 
The maximal allowed $\phi_{\mu}$ is roughly inversely proportional to $\tan\beta$, e.g.  for $\tan\beta=10$
we have $a=0.005$ (Fig.\ref{tanb10}). 
In the case of  the parton  model, the 
cancellations occur, for example, with $a=0.006$ and 
$\tan\beta=3$, $m_0=m_{1/2}=200$ GeV, and $\vert A \vert=20$ GeV
(Fig.\ref{parton}). As mentioned earlier, the cancellations are harder to achieve in the parton model,
which results in tighter limits on $\phi_{\mu}$ ($\sim{\cal O}(10^{-3})$).
In both cases $\phi_{\mu}$ is restricted to be of order $10^{-3}-10^{-2}$,
whereas $\phi_A$ can be arbitrary (however  physical effects due to $\phi_A$ will be suppressed
because of the small magnitude of $A$). The GUT parameters given above imply that the squark masses
at low energies are about 500 GeV, so these bounds can be relaxed only if the squark masses are over 1 TeV. 

\begin{figure}[ht]
\begin{center}
\begin{tabular}{c}
\epsfig{file=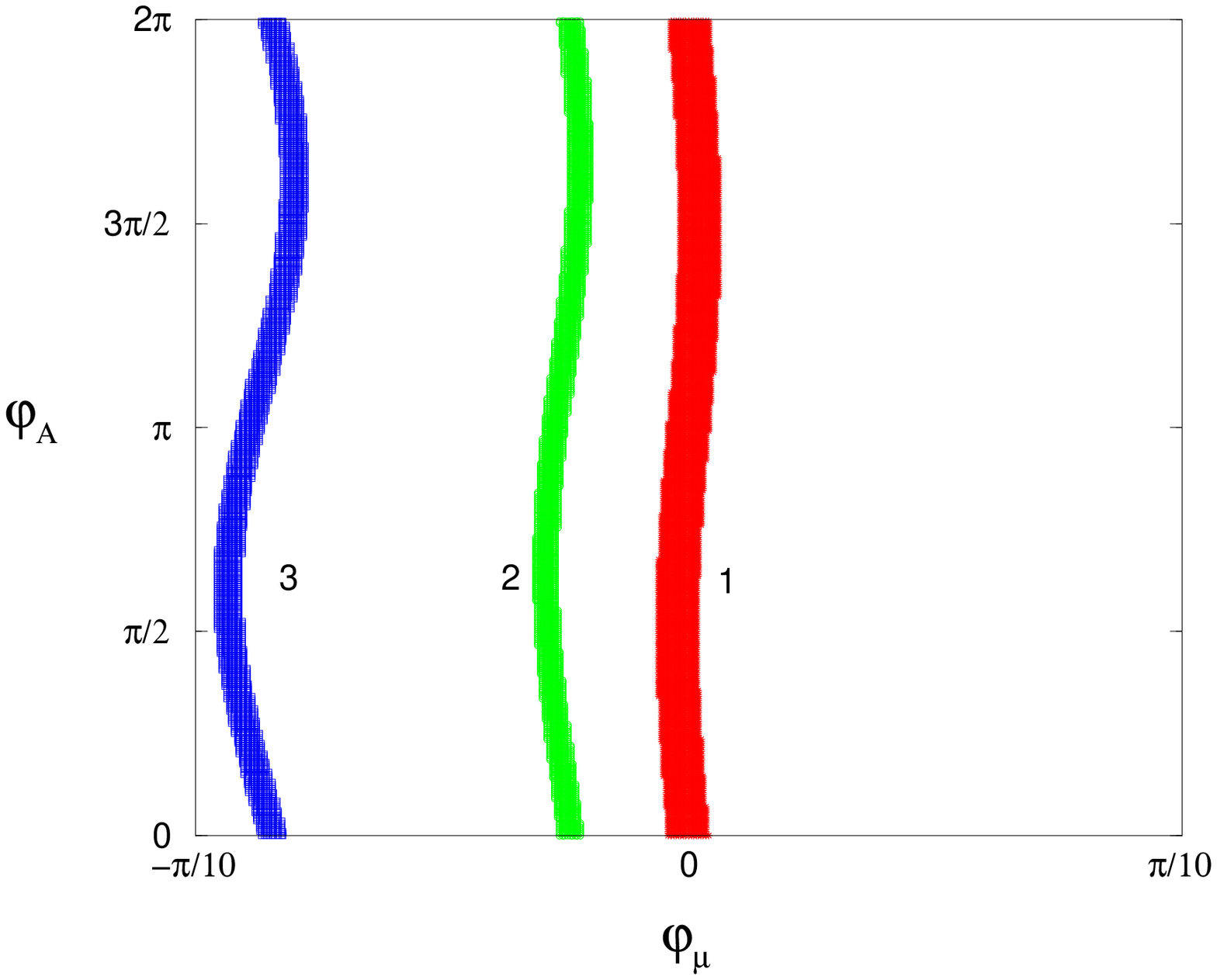, width=11cm, height=8.0cm}\\  
\end{tabular}
\end{center}
\caption{ Bands allowed by  the electron (1), neutron (2), and mercury
(3) EDMs
cancellations in the mSUGRA-type model with a nonzero gluino phase. 
Here  $\tan\beta=3$, $m_0=m_{1/2}=200$ GeV, $A=40$ GeV, $\phi_3=\pi/10$. }
\label{phi3}
\end{figure}

\begin{figure}[ht]
\begin{center}
\begin{tabular}{c}
\epsfig{file=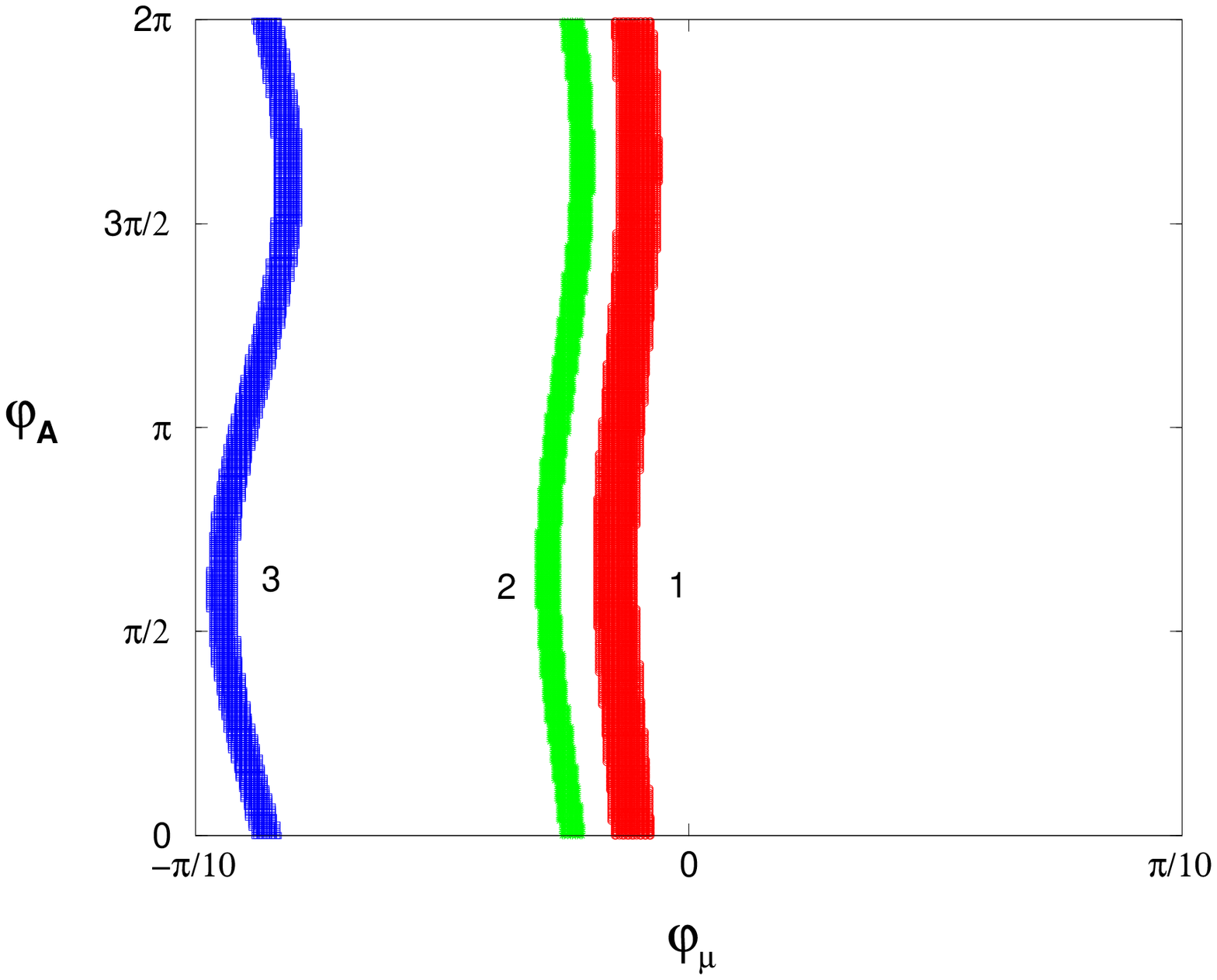, width=11cm, height=8.0cm}\\  
\end{tabular}
\end{center}
\caption{ Bands allowed by  the electron (1), neutron (2), and mercury
(3) EDMs
cancellations in the mSUGRA-type model with  nonzero gluino and bino  phases. 
Here  $\tan\beta=3$, $m_0=m_{1/2}=200$ GeV, $A=40$ GeV, $\phi_1=\phi_3=\pi/10$. }
\label{phi1phi3}
\end{figure}

If the gluino phase is turned on, simultaneous EEDM, NEDM, and mercury EDM  cancellations are not 
possible. The gluino phase affects the NEDM cancellation band by altering the relation
(\ref{band}):
\begin{equation}
\phi_{\mu} \simeq - a \sin (\phi_A +\alpha) - c \;,
\label{band1}
\end{equation}
while leaving the EEDM cancellation band almost intact (Fig.\ref{phi3}). 
An introduction of the bino phase $\phi_1$ qualitatively has the same ``off-setting'' effect on
the EEDM cancellation band as the gluino phase does on that of the NEDM (Fig.\ref{phi1phi3}).  
Note that the bino phase has no significant effect on the neutron and mercury cancellation bands 
since the neutralino contribution in both cases is small. 
When both the gluino
and bino phases are present (and fixed), simultaneous electron, neutron, and mercury EDMs cancellations
do not appear to be possible along a band. The reason is that the mercury EDM depends
on $\phi_3$ more strongly than the NEDM does, so that an introduction of the gluino phase splits
the bands where the mercury and neutron EDM cancellations occur as illustrated in Figs.\ref{phi3} and
\ref{phi1phi3}. Thus we see that zero gaugino phases are much more preferred by the cancellation
scenario.

The cancellation scenario involves a significant fine-tuning. Indeed, restricting
the phases to the band where the cancellations occur does not increase the symmetry
of the model and thus is unnatural according to the t'Hooft's criterion \cite{thooft}.
It is a non-trivial task to quantify the degree of fine-tuning in this case.
One possibility is to define an EDM sensitivity coefficient with respect to the
CP-phases, in analogy with Eq.\ref{sens}. 
We typically find it to be between 30 and 100 on the cancellation band. 
This  however represents only
a local behaviour of the EDMs. In other words it shows how easy it is to
spoil the cancellations without a reference to how improbable it is to achieve such
cancellations in the first place. An alternative way to quantify the degree of
fine-tuning is simply to estimate the probability that a random small area in the 
$(\phi_{\mu},\phi_A)$ plane will safisfy the cancellation condition. Since any point
is not prefered over any other point by the underlying theory, this should give
a fairly good idea of the $minimal$ degree of fine-tuning needed. From Fig.\ref{sugra} we obtain
\begin{equation}
{\rm fine-tuning}\; \sim \;  ({\rm probability})^{-1} \; \gsim \; 10^2\;.
\end{equation}
Note that this estimate does not take into account the fine-tuning of the other soft 
breaking parameters which is necessary to allow for simultaneous EEDM, NEDM, 
and mercury EDM cancellations. Other estimates give a similar number for the universal case, 
whereas for the nonuniversal case the degree of fine-tuning drastically $increases$:
only one out of $10^5$ random points in the parameter space satisfies the cancellation
condition \cite{barger}.

\subsubsection{EDM cancellations in D-brane models.}

Let us first briefly review basic ideas of D-brane models (see also Refs.\cite{Ibanez:1999rf}
and \cite{carlos}).
Recent studies of type I strings have shown that it is possible to construct a number of 
models with non--universal soft SUSY breaking terms which are phenomenologically 
interesting. Type I models can contain $9$-branes, $5_i$-branes, $7_i$-branes, and 
$3$-branes where the index $i=1,2,3$ denotes the complex compact coordinate which is
included in the $5$-brane world volume or which  is orthogonal  to the $7$-brane world 
volume. However, to preserve $N=1$ supersymmetry in $D=4$ not all of these branes 
can  be present simultaneously and we can have (at most) either D9-branes with 
D$5_i$-branes or D3-branes with D$7_i$-branes.

Gauge symmetry groups are associated with stacks of branes located ``on top of each other''.
A stack of $N$ branes corresponds to the group $U(N)$. The matter fields are
associated with open strings which start and end on the branes. These strings may be attached
to either the same stack of branes or two different sets of branes which have
overlapping world volumes. The ends of the string carry quantum numbers associated
with the symmetry groups of the branes.
For example, the quark fields have to be  attached to the $U(3)$ set of branes,
while the  quark doublet  fields also have to be attached to the $U(2)$ set of branes.
Given a brane configuration, the Standard Model fields are constructed according to their
quantum numbers.

\begin{figure}[ht]
\begin{center}
\begin{tabular}{c}
\epsfig{file= 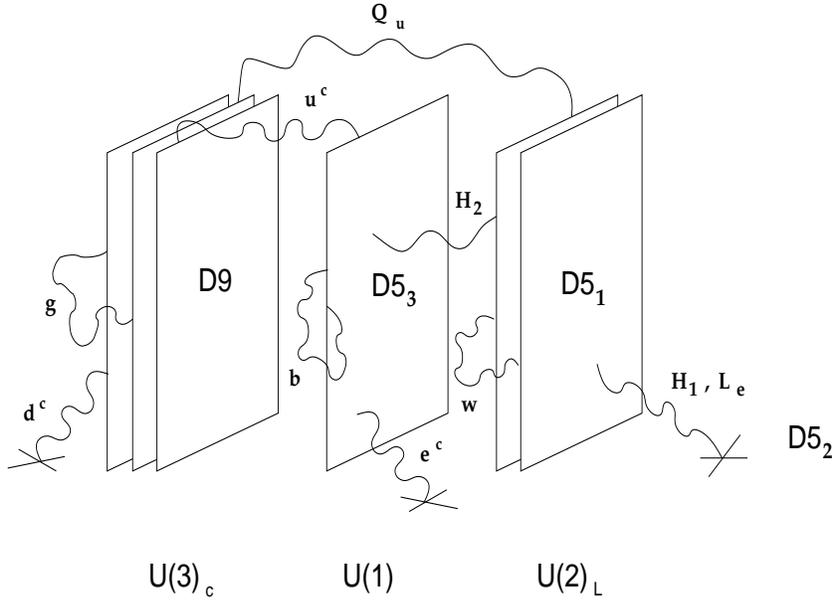, width=11cm, height=8.0cm}\\  
\end{tabular}
\end{center}
\caption{Embedding the SM gauge group within  different sets of D-branes. The extra
$D_q$ brane ($5_2$) is marked by a cross. }
\label{branes}
\end{figure}

The SM gauge group can be obtained in the context of D-brane scenarios from 
$U(3)\times U(2)\times U(1)$, where the $U(3)$ arises from three coincident branes, 
$U(2)$ arises from two coincident D-branes and $U(1)$ from one D-brane. As explained
in detail in Ref.\cite{carlos},  there are different possibilities for 
embedding the SM gauge groups within these D-branes. It was shown that if the SM gauge 
groups come from the same set of D-branes, one cannot produce the correct values 
for  the gauge couplings $\alpha_j(M_Z)$ and the presence of additional matter (doublets
and triplets) is necessary to obtain the experimental values of the couplings 
\cite{Bailin:2000kd}. 
On the other hand, the assumption  that the SM 
gauge groups originate from different sets of $D$-branes leads in a natural 
way to intermediate values for the string scale $M_S \simeq 10^{10-12}$ GeV
\cite{carlos}. 
In this case, the analysis of the soft terms has been done under the 
assumption that only the dilaton and moduli fields contribute to supersymmetry breaking
and it has been found  that these soft terms  are generically non--universal. 
The MSSM fields arising from  open strings   are shown in Fig.\ref{branes}.
For example, the up quark singlets $u^c$ are states of the type $C^{9 5_3}$,
the quark doublets are $C^{9 5_1}$, etc.
The presence of extra ($D_q$) branes which are not associated with the SM gauge groups
is often necessary to reproduce the correct hypercharge and to cancel non-vanishing
tadpoles.

Recently there has been a considerable interest in supersymmetric models 
derived from D-branes \cite{lisa},\cite{brane}. 
In a toy model of Ref.\cite{lisa},  the gauge group
$SU(3)_c \times U(1)_Y$ was associated with  $5_1$ branes and  $SU(2)_L$ was associated with
$5_2$ branes. It was shown that in this model the gaugino masses are non--universal
($M_1=M_3 \ne M_2$) so that the physical CP phases 
 are $\phi_1 = \phi_3$, $\phi_A$ and $\phi_{\mu}$. Non-universality of the gaugino
masses allowed to enlarge the regions of the  parameter space where the NEDM/EEDM
cancellations occured. However, such a model is oversimplified and 
 the relation $\phi_1 = \phi_3 \not = \phi_2$  does not appear to  hold in 
more realistic models \cite{us}.
In what follows we will consider the EDM cancellations in this and  a more realistic D-brane models.

The model in which $U(3)$, $U(2)$, and $U(1)$ originate from different sets of branes
is phenomenologically interesting.
 In this case one naturally obtains an
intermediate string scale ($10^{10}-10^{12}$ GeV), although  higher values
up to $10^{16}$ GeV are still allowed. 
Both the up and the down type Yukawa interactions are allowed, while that for
the leptons typically vanishes (depending on further details of the model) \cite{carlos}.
The (anomaly-free) hypercharge is expressed in terms of the $U(1)$ charges $Q_{1,2,3}$ of the 
$U(1)_{1,2,3}$ groups:
\begin{equation}
Y=-{1\over 3} Q_3 -{1\over 2} Q_2 +Q_1\;,
\end{equation}
with the following $(Q_3,Q_2,Q_1)$ charge assignment:
\begin{eqnarray}
&& q=(1,-1,0) \;, \; u^c=(-1,0,-1) \;,\; d^c=(-1,0,0)\;,\\ \nonumber
&& l=(0,1,0) \;, \; e^c=(0,0,1) \;,\;\\ \nonumber
&& H_2=(0,1,1)\;,\; H_1=(0,1,0) \;.
\end{eqnarray}

The gaugino masses in this model are given by 
\begin{eqnarray}
M_3 & = & \sqrt{3} m_{3/2} \sin \theta \; e^{-i\alpha_s} \ ,  \\
M_{2} & = & \sqrt{3}  m_{3/2}\ \Theta_1 \cos \theta\; e^{-i\alpha_1} \ , \nn\\
M_{Y} & = &  \sqrt{3}  m_{3/2}\ \alpha_Y (M_S) \nonumber \\ 
      & \times & \left(\frac{2}{\alpha_1 (M_S)}\Theta_3 \cos \theta e^{-i\alpha_3}
+\frac{1}{\alpha_2 (M_S)}\Theta_1 \cos \theta e^{-i\alpha_1}
+\frac{2}{3\alpha_3 (M_S)}\sin \theta e^{-i\alpha_s}
\right) \;, \nonumber
\label{gaugino1}
\end{eqnarray}
where 
\begin{equation}
\frac{1}{\alpha_Y(M_S)} =
\frac{2}{\alpha_1(M_S)} + \frac{1}{\alpha_2(M_S)}
+ \frac{2}{3 \alpha_3(M_S)}\ .
\label{couplings}
\end{equation} 
Here $\alpha_k$ correspond to the gauge couplings of the $U(k)$ branes.
As  shown in Ref.\cite{carlos},  $\alpha_1(M_S) \simeq 0.1$ leads 
to the string scale $M_S \approx 10^{12}$ GeV. The parameters $\theta$ and 
$\Theta_i$ parameterize supersymmetry breaking in the usual way \cite{Ibanez:1999rf}:
\begin{eqnarray}
&& F^S= \sqrt{3} (S+S^*) m_{3/2} \sin \theta \;e^{-i\alpha_s} \;, \nonumber \\
&& F^i= \sqrt{3} (T_i+T^*_i) m_{3/2} \cos \theta\; \Theta_i e^{-i\alpha_i} \;,
\end{eqnarray}
and
$i=1,2,3$ labels the three complex compact dimensions.

The soft scalar masses are given by
\begin{eqnarray}
m^2_{q} & = & m_{3/2}^2\left[1 -
\frac{3}{2}  \left(1 - \Theta_{1}^2 \right)
\cos^2 \theta \right] \ , \nn \\
m^2_{d^c} & = & m_{3/2}^2\left[1 -
\frac{3}{2}  \left(1 - \Theta_{2}^2 \right)
\cos^2 \theta \right] \ , \nn \\
m^2_{u^c} & = & m_{3/2}^2\left[1 -
\frac{3}{2}  \left(1 - \Theta_{3}^2 \right)
\cos^2 \theta \right] \ , \nn \\
m^2_{e^c} & = & m_{3/2}^2\left[1- \frac{3}{2}
\left(\sin^2\theta + \Theta_{1}^2 \cos^2\theta  \right)\right] \ , \nn \\
m^2_{l} & = & m_{3/2}^2\left[1- \frac{3}{2}
\left(\sin^2\theta + \Theta_{3}^2 \cos^2\theta  \right)\right] \ , \nn \\
m^2_{H_2} & = & m_{3/2}^2\left[1- \frac{3}{2}
\left(\sin^2\theta + \Theta_{2}^2 \cos^2\theta  \right)\right] \ , \nn \\
m^2_{H_1} & = & m^2_l \;,
\label{scalars1}
\end{eqnarray}
and the trilinear parameters are
\begin{eqnarray}
A_{u} & = &  \frac{\sqrt 3}{2}m_{3/2}
   \left[\left(\Theta_{2} e^{-i\alpha_2} - \Theta_1 e^{-i\alpha_1} 
 - \Theta _{3} e^{-i\alpha_3}  \right) \cos\theta
- \sin\theta \;e^{-i\alpha_s} \right] \ ,
\label{trintrin}
\\
A_{d} & = &  \frac{\sqrt 3}{2}m_{3/2}
   \left[\left(\Theta_{3} e^{-i\alpha_3} - \Theta_1 e^{-i\alpha_1} 
  - \Theta _{2} e^{-i\alpha_2}  \right) \cos\theta
- \sin\theta \;e^{-i\alpha_s} \right] \ ,
\label{trintrintrin}
\\
A_{e} & = &  0\; .
\label{trilin11}
\end{eqnarray}

We observe that the angles $\Theta_i$ and $\theta$ are quite constrained if we are to avoid
negative mass-squared's for squarks and sleptons.
In what follows  we set  $\Theta_3=0$ and $\alpha_1=\alpha_2$; 
then the soft terms are parameterized in terms of the phase $\phi \equiv \alpha_1-\alpha_s$.

\begin{figure}[ht]
\begin{center}
\begin{tabular}{c}
\epsfig{file=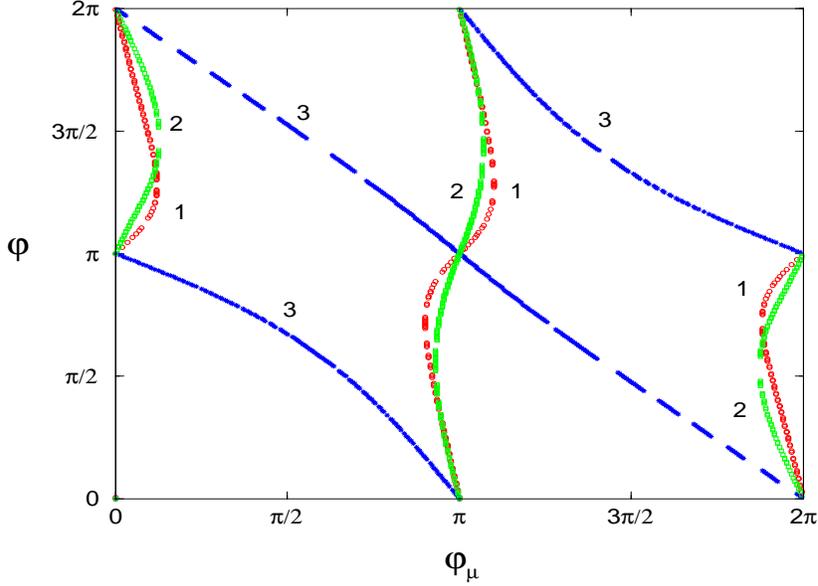, width=11cm, height=8.0cm}\\  
\end{tabular}
\end{center}
\caption{Bands allowed by  the electron (1), neutron (2), and mercury
(3) EDMs
 in the D-brane model.
Here  $\tan\beta=3$, $m_{3/2}=150$ GeV, $\Theta_1^2=\Theta_2^2=1/2$,
$\cos^2\theta=2 \sin^2\theta=2/3$, and $M_S \sim 10^{16}$ GeV. }
\label{branecancel1}
\end{figure}

\begin{figure}[ht]
\begin{center}
\begin{tabular}{c}
\epsfig{file=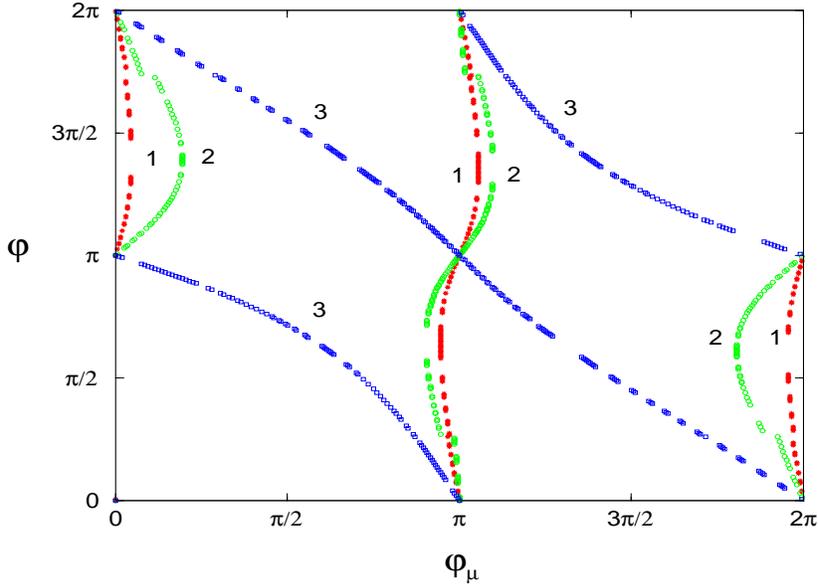, width=11cm, height=8.0cm}\\  
\end{tabular}
\end{center}
\caption{Bands allowed by  the electron (1), neutron (2), and mercury
(3) EDMs
 in the D-brane model with an intermediate scale.
Here  $\tan\beta=3$, $m_{3/2}=150$ GeV, $\Theta_1^2=\Theta_2^2=1/2$,
$\cos^2\theta=2 \sin^2\theta=2/3$, and $M_S \sim 10^{12}$ GeV. }
\label{intermediate}
\end{figure}

In Fig.\ref{branecancel1} we display the bands allowed by the electron (red), neutron (green), and
mercury (blue) EDMs. 
In this figure,
we set $m_{3/2}=150$ GeV, $\tan\beta=3$, $\Theta_1^2=\Theta_2^2=1/2$, 
$\cos^2\theta=2\sin^2\theta=2/3$,
and $\alpha_1(M_S) \sim 1$ with $M_S$ being the GUT scale.
For the plot to be more illustrative, we do not impose any additional constraints
besides the EDM ones (i.e. bounds on the chargino and  slepton masses, etc.).
It is clear that even though simultaneous EEDM/NEDM cancellations allow 
the phase $\phi$ to be  ${\cal{O}}(1)$, an addition of the  mercury constraint
 requires all phases to be very small (modulo $\pi$) and thus
practically rules out the cancellation scenario in this context.
The situation becomes even worse in the case of an intermediate string scale $\sim 10^{12}$ GeV
(i.e. $\alpha_1(M_S) \sim 0.1$), see Fig.\ref{intermediate}.
We find it quite generic  that the mercury
EDM behaviour in D-brane models  is very different from that of the electron and neutron
and thus is crucial in constraining the parameter space. The major difference from
the mSUGRA-type models with fixed $\phi_Y$ and $\phi_3$ is that the phase of the A-terms
is correlated with the gaugino phases resulting in the cancellation bands which are
not described by a simple relation  
$\phi_{\mu} \simeq - a \sin (\phi_A +\alpha) - c$.

\begin{figure}[ht]
\begin{center}
\begin{tabular}{c}
\epsfig{file=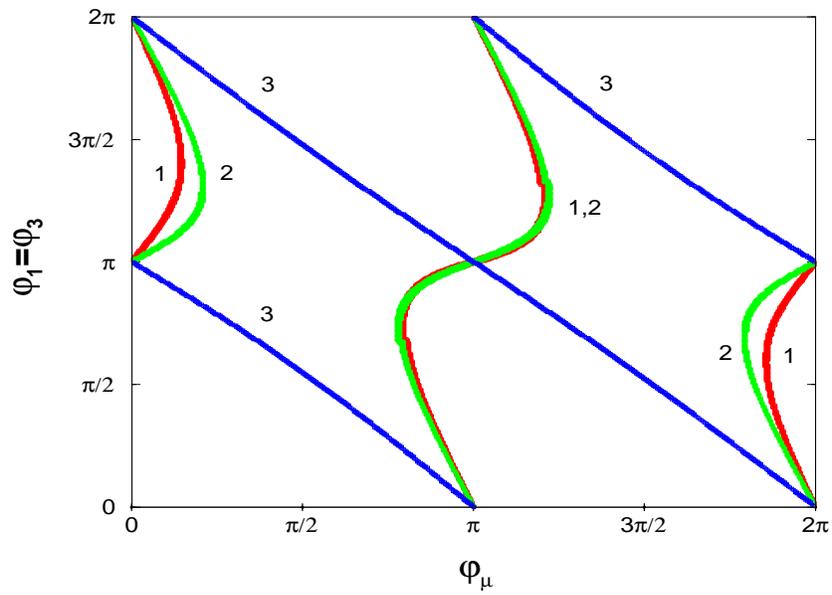, width=11cm, height=8.0cm}\\  
\end{tabular}      
\end{center}
\caption{Bands allowed by  the electron (1), neutron (2), and mercury
(3) EDMs
 in the model of Ref.\cite{lisa}.
Here  $\tan\beta=2$, $m_{3/2}=150$ GeV, $\Theta_1=0.9$,
$\theta=0.4$. }
\label{branecancel2}
\end{figure}

Next we consider the model of Ref.\cite{lisa}. 
The (corrected) soft terms for this model read (for $\Theta_3=0$)
\begin{eqnarray}
&& M_Y=M_3=-A=\sqrt{3}m_{3/2} \cos\theta\; \Theta_1 e^{-i \alpha_1}\;, \nonumber\\
&& M_2=\sqrt{3} m_{3/2} \cos\theta\; \Theta_2 e^{-i \alpha_2}\;,\nonumber\\
&& m^2_L= m_{3/2}^2 (1-{3\over 2} \sin^2\theta);\,\nonumber\\
&& m^2_R= m_{3/2}^2 (1-3 \cos^2\theta \;\Theta_2^2)\;.
\end{eqnarray}
To illustrate the EDM constraints, 
we choose the parameters which allow for simultaneous EEDM/NEDM cancellations,
namely  $m_{3/2}=150$ GeV, $\tan\beta=2$, $\Theta_1=0.9$, 
and $\theta=0.4$ as given in Ref.\cite{lisa}. Fig.\ref{branecancel2} shows that the mercury constraint has the
same behaviour as in the model considered above and rules out large CP-phases.

We see that the cancellation scenario in simple models faces a number of difficulties.
Presently available string-motivated models with non-universal gaugino masses cannot
accommodate simultaneous electron, neutron, and mercury EDM cancellations.
In the mSUGRA, such cancellations are possible but require a significant fine-tuning.
The addition of the mercury EDM bound restricts the phase of the $\mu$ term to be 
${\cal O}(10^{-2})$ if we are to achieve the EDM cancellations along a band in the $(\phi_A, \phi_{\mu})$
plane.  Without an additional SUSY flavour structure, the CP-phases
allowed by the cancellations will have very small observable  effects. 
 Even in a more general situation
(unconstrained MSSM), 
the phases allowed by the EDM cancellations 
typically lead to small CP-asymmetries ($\lsim 1\%$) in collider experiments \cite{barger}. 
Testing the cancellation scenario experimentally may prove to be a challenge.

\subsection{ Flavour-off-diagonal CP violation}

This is one of the more attractive possibilities to avoid overproduction of EDMs in SUSY models.
Nonobservation of EDMs may imply that CP-violation  has a flavour-off-diagonal character
just like in the Standard Model. 
The origin of CP-violation in this case is closely related to the origin of the flavour structures
rather than the origin of supersymmetry breaking. 
While models with flavour-off-diagonal CP violation
naturally avoid the EDM problem, they have testable  effects in  K and B physics.

This class of models requires hermitian Yukawa matrices and A-terms, which forces the flavour-diagonal phases
to vanish (up to small RG corrections) in any basis. The flavour-independent quantities such as
the $\mu$-term, gaugino masses, etc. are real.
This is naturally implemented  in left-right symmetric models \cite{mohapatra} and models with
a horizontal flavour symmetry \cite{Abel:2000hn}. 

In the left-right models, the hermiticity of the Yukawas and A-terms as well as the reality 
of the $\mu$-term  is forced by the left-right symmetry. The $SU(2)_L$ gaugino mass is in general complex,
so in order to suppress the EDMs the additional assumption of its reality is needed. The phenomenology
of such models in the context of up-down unification has been studied in \cite{mohapatra1}. The left-right
symmetry appears to be too restrictive in this case to satisfy all of the phenomenological constraints;
however the decisive test will be undertaken at the B factories.

Another possibility is based on a horizontal $U(3)_H$ symmetry \cite{Abel:2000hn}.  
Hermitian Yukawa matrices may appear due to a (gauged) horizontal symmetry $U(3)_H$ which gets
broken spontaneously by the VEVs of the  real adjoint fields $T^a_{\alpha}$
($a=1..9;\; \alpha=u,d,l,..)$ \cite{masiero}.  Some of these {\em real} VEV's also break
CP since some of the components of $T^a_{\alpha}$ are CP-odd.  As a
result, CP violation appears in the superpotential through complex
Yukawa couplings, whereas the $\mu$-term is real since it arises from a $U(3)_H$
invariant combination of the type $T^a_{\alpha}T^a_{\beta}$.
 An effective  $U(3)_H$-invariant superpotential of the type
$ \hat W = {g_H\over M}  \hat H_1 \hat Q_i ( T^a_d\lambda^a)_{ij} \hat D_j$
produces the Yukawa matrix 
$(Y^d)_{ij}= {g_H\over M} \langle T^a_d \rangle (\lambda^a)_{ij}$, where $\lambda^0$
is proportional to the unit matrix  and $\lambda^{1-8}$ are the Gell-Mann matrices.
Note that the $real$ fields  $T^a_{\alpha}$ may only come from a non-supersymmetric (anti-brane) sector.
If $T^a_{\alpha}$ are the scalar components of chiral multiplets, they are intrinsically
complex and hermitian Yukawas in this case arise only if the VEVs of $T^a_{\alpha}$ are real. 

The gaugino masses in this model are not forced to be real by symmetry. One has to make an additional
assumption that either the gaugino masses are universal and the corresponding phase 
can be rotated away or that the SUSY breaking dynamics conserve CP which seems natural if CP breaking 
is associated
with the origin of flavour structures. In other words,  the SUSY breaking auxiliary fields get real VEVs
as a result of the underlying dynamics such as the dilaton stabilization in Type I string models
\cite{Abel:2001tf}      
or the effective potential minimization in heterotic string models \cite{Bailin:1998yt}.
 Further, if the K\"ahler potential is either generation
independent or left-right symmetric, the A-terms are hermitian \cite{Abel:2000hn}. 
This leads to very small phases in the flavour-diagonal mass insertions which are responsible for generating
EDMs and the EDM bounds are easily satisfied.

Both the left-right model and the model with a horizontal symmetry mitigate the strong CP-problem
(under the additional assumptions given above). The $\bar \theta$ parameter vanishes at both the tree 
and the leading log one-loop levels. However neither of these models solves the problem completely due to 
the significant one loop finite corrections which appear mostly due to a non-degeneracy of the
squark masses  and a misalignment between the Yukawas and the A-terms \cite{Dine:1993qm}.

In this class of models we have the following setting at the high energy scale:
\begin{eqnarray}
&& Y_{\alpha}=Y_{\alpha}^{\dagger}\;, A_{\alpha}=A_{\alpha}^{\dagger} \;, \nonumber\\
&& {\rm Arg} (M_k)= {\rm Arg} (\mu)=0\;.
\end{eqnarray}
Generally, the off-diagonal elements of the A-terms can have ${\cal{O}}(1)$ phases
without violating the EDM constraints.
Due to the RG effects,  large phases in the soft
trilinear couplings involving the third generation generate small phases in the
flavour-diagonal mass insertions for the light generations, and thus induce the EDMs.
For example, the A-terms  of the form\footnote{The standard supergravity relation 
$(\hat A_{\alpha})_{ij}=(A_{\alpha})_{ij} (Y_{\alpha})_{ij}$ for the 
trilinear soft couplings is assumed.}
\begin{equation}
A_d=A_u=m_0  \left( \matrix{1 & a_{12} & a_{13} \cr
                     a_{12}^* & 1 & a_{23} \cr
                     a_{13}^* & a_{23}^* & a_{33}} \right) \;
\end{equation}
and  the following GUT-scale hermitian Yukawa matrices  
\begin{eqnarray}
&& Y^u=\left( \matrix {4.1\times 10^{-4}&6.9\times 10^{-4}\;{\rm i}&-1.4\times 10^{-2} \cr
-6.9\times 10^{-4}\;{\rm i}&3.5\times 10^{-3}&-1.4\times 10^{-5}\;{\rm i}\cr
-1.4\times 10^{-2}&1.4\times 10^{-5}\;{\rm i}&6.9\times 10^{-1}
             } \right)\; , \nonumber \\
&& Y^d=\left( \matrix {1.3\times 10^{-4}&2.0\times 10^{-4}+1.8\times 10^{-4}\;{\rm i}&-4.4\times 10^{-4} \cr
2.0\times 10^{-4}-1.8\times 10^{-4}\;{\rm i}&9.3\times 10^{-4}&
7.0\times 10^{-4}\;{\rm i}\cr
-4.4\times 10^{-4} & -7.0\times 10^{-4}\;{\rm i} & 1.9\times 10^{-2}
             } \right)\; . \nonumber
\label{yukawa}
\end{eqnarray}
typically induce the mercury EDM of the order of the experimental limit if $a_{33}$ is of order 1, 
whereas the induced NEDM is 1-2 orders of magnitude below the experimental limit. If one uses 
Yukawa textures with smaller $Y^{\alpha}_{13}$, this RG effect will be suppressed  rendering
the induced mercury EDM far below the experimental limit. 

The supersymmetric contribution to the $\varepsilon'$ parameter is suppressed in models with
hermitian flavour structures\footnote{For phenomenology of models with non-universal 
(but non-hermitian) A-terms see also \cite{nonuniversal}.} \cite{Abel:2000hn}. 
This occurs due to severe cancellations between the
contributions involving $(\delta_{12}^d)_{LR}$ and
$(\delta_{12}^d)_{RL}$ mass insertions (we use the standard
definitions of \cite{Gabbiani:1996hi}). Due to the hermiticity
$(\delta_{12}^d)_{LR} \simeq (\delta_{12}^d)_{RL}$, whereas they
contribute to $\varepsilon'$ with opposite signs. Typically we
find $\Bigl\vert {\rm Im} \left[(\delta_{12}^d)_{LR}- (\delta_{12}^d)_{RL}
\right]\Bigr\vert < 10^{-6}$ which produces  $\varepsilon'$ an order
of magnitude below the experimental limit \cite{Gabbiani:1996hi}.
On the other hand, similar cancellations do not occur for the $\varepsilon$
parameter  and the SUSY contribution to $\varepsilon$ can be  even dominant.
The value of  $(\delta_{12}^d)_{LR}$  which saturates the 
observed  $\vert  \varepsilon \vert \simeq 2.26 \times 10^{-3} $ is given by $\sqrt{\vert
\mathrm{Im}(\delta_{12}^d)^2_{LR} \vert } \simeq 3.5 \times 10^{-4}$
for the gluino and squark masses of  500 GeV \cite{Gabbiani:1996hi}.

For completeness, we provide the bounds on the imaginary parts of the mass insertions
\begin{equation}
(\delta_{ii}^{d(u)})_{LR}= {1\over \tilde m^2}
 ( (\hat A^{d(u) \dagger}_{SCKM})_{ii} v_{1(2)} - Y_i^{d(u)} \mu
v_{2(1)}) \;
\end{equation}
derived from the leading gluino contributions to the electromagnetic (NEDM) operator  
 and the bino contribution to the electron EDM. 
We update the bounds of Ref.\cite{Gabbiani:1996hi}
and also include the QCD correction factor. The advantage of the mass insertion approach is
that it allows to obtain model independent bounds and thus is quite useful when
dealing with complicated flavour structures. For comparison we present the bounds
for the chiral (Table 1) and the parton (Table 2) neutron models. 

A drastic improvement of the bounds comes from the addition of the mercury EDM constraint.
Using the expressions for the chromomagnetic moments of Ref.\cite{savoy},
 in Table 3 we present  the  bounds on the mass insertions
from the gluino contributions to  the mercury EDM. 
For $\vert {\rm Im}(\delta_{11}^{d})_{LR} )  \vert$ and
$\vert {\rm Im}(\delta_{11}^{u})_{LR} )  \vert$  these bounds turn out to be very strict,
more than an order of magnitude stricter than those imposed by the NEDM. 
This severely restricts
the CP-asymmetry $A_{CP}(b\rightarrow s \gamma)$ in models with hermitian flavour structures.
The reason is that in order to obtain 
a large ($\sim 10\%$) SUSY contribution to this observable, the elements of the A-terms 
involving the third generation have to be larger than 1.  This induces 
via the RG running a 
considerable $\vert {\rm Im}(\delta_{11}^{d,u})_{LR} )  \vert$, often in conflict with
the bounds of Table 3. 
We find that with the above Yukawa textures  
$A_{CP}(b\rightarrow s \gamma)$ is allowed to be no more than  $2$-$3\%$.
One can relax this constraint by using different hermitian textures, especially
with suppressed $Y^{\alpha}_{13}$. In this case the CP-asymmetry can be as large
as $6$-$7\%$. 

To conclude this section, we remark that the problem of baryogenesis in this class of models
requires careful investigation and at the moment it is unclear whether or not large flavour
off-diagonal SUSY phases can produce sufficient baryon asymmetry of the universe.

\section{Discussion and conclusions}

We have systematically analyzed  constraints on supersymmetric models imposed by the 
experimental bounds on the electron, neutron, and mercury electric dipole moments.
We find that the EDMs can be suppressed in SUSY models with 

1) small SUSY CP-phases ($\lsim 10^{-2}$). This possibility can be motivated by the 
approximate CP-symmetry which also implies that the CKM phase is small. This
provides testable signatures for  B-factories' experiments.

2) heavy SUSY scalars, $m_{sfermion} \sim 10$ TeV. In this class of models there is
a large hierarchy between the SUSY and electroweak scales, which is hard to 
realize without an extreme fine-tuning. 

3) EDM cancellations. We have analyzed the possibility of such cancellations 
in D-brane and mSUGRA-like models with nontrivial gaugino phases. We find that,
with the addition of the mercury EDM constraint, only the EDM cancellations in mSUGRA
($\phi_1 \simeq \phi_3 \simeq 0$) survive in any considerable part of the parameter space.
Even in this case the cancellations require small $\phi_{\mu} \sim 10^{-2}$ and 
suppressed $\vert A \vert$ ($\sim 0.1 m_0$). As a result, the border between the 
small phases and the cancellation scenarios fades away. In addition to the finetuning
problem,  models with the EDM cancellations lack predictive power as it is unclear
whether the allowed CP-phases can have observable effects.

4) flavour-off-diagonal CP violation. This can occur in models with CP-conserving
SUSY breaking dynamics and hermitian flavour structures. Such  models
allow for ${\cal O}(1)$ flavour off-diagonal phases which can have significant
effects in  K and B physics.

It is also possible to combine different mechanisms to suppress the EDMs. For example,
a ``hybrid'' of the decoupling and the cancellation scenarios was considered in 
Ref.\cite{Chattopadhyay:2000fj}. Such models seem to share shortcomings of
both ``parents'' without an apparent advantage over either of them. 

There is also a rather radical proposal that all gaugino masses 
and the A-terms are vanishingly small \cite{Clavelli:2000ua}.
Clearly this eliminates all of the physical phases in Eq.(\ref{phases}) and thus produces
no CP violation. 
Such a strong assumption requires a firm  motivation such as  the continuous R-symmetry.
However, by the same token, the R-symmetry eliminates the $B\mu$ term thereby leading to
a very light CP-odd Higgs boson.
Such  a scenario faces a number of difficulties with experimental results.

To summarize, we have studied the supersymmetric CP problem taking into account all of the 
current EDM constraints. Our conclusion is that there remain  two attractive ways to avoid
overproduction of the EDMs in SUSY models. The first possibility is that CP is an approximate
symmetry of nature and the second one is that CP violation  has a flavour off-diagonal character 
just as in the Standard Model.\\

{\bf Acknowledgements.}  S.A. and O.L. were supported by the PPARC Opportunity grant, S.K. was
supported by the PPARC SPG.
The authors are grateful to D. Bailin, M. Brhlik, T. Ibrahim,  
C. Mu\~{n}oz, and M. Pospelov for useful discussions, and to M. Gomez and D. Cerde\~{n}o 
for their help in the computing aspects  of this work.

\newpage

\begin{table}
\begin{center}
\begin{tabular}{|c||c|c|c|}
\hline
  $x$   & $\vert {\rm Im}(\delta_{11}^{d})_{LR} )  \vert$ & $\vert {\rm Im}(\delta_{11}^{u})_{LR} )  \vert$ &
$\vert {\rm Im}(\delta_{11}^{l})_{LR} )  \vert$ \\
\hline
\hline
0.1 & $1.0\times 10^{-6}$ & $2.0\times 10^{-6}$ & $1.4\times 10^{-7}$ \\
0.3 & $9.2\times 10^{-7}$ & $1.8\times 10^{-6}$ & $1.3\times 10^{-7}$ \\
1 & $1.1\times 10^{-6}$ & $2.2\times 10^{-6}$ & $1.6\times 10^{-7}$ \\
3 & $1.8\times 10^{-6}$ & $3.6\times 10^{-6}$ & $2.6\times 10^{-7}$ \\
5 & $2.4\times 10^{-6}$ & $4.9\times 10^{-6}$ & $3.5\times 10^{-7}$ \\
10 & $4.1\times 10^{-6}$ & $8.2\times 10^{-6}$ & $5.9\times 10^{-7}$ \\
\hline
\end{tabular}
\end{center}
\caption{Bounds on the imaginary parts of the mass insertions. The chiral quark 
model for the neutron is assumed.
 Here $x=m_{\tilde g}^2/m_{\tilde q}^2=
m_{\tilde B}^2/m_{\tilde l}^2$, $ m_{\tilde q}=500 \;GeV, m_{\tilde l}=100\; GeV$. For different 
squark/slepton masses the bounds are to be multiplied by $m_{\tilde q}/500\; GeV$ or
$m_{\tilde l}/100 \;GeV$.}
\end{table}

\begin{table}
\begin{center}
\begin{tabular}{|c||c|c|c|}
\hline
  $x$   & $\vert {\rm Im}(\delta_{11}^{d})_{LR} )  \vert$ & $\vert {\rm Im}(\delta_{11}^{u})_{LR} )  \vert$ &
$\vert {\rm Im}(\delta_{22}^{d})_{LR} )  \vert$ \\
\hline
\hline
0.1 & $1.8\times 10^{-6}$ & $1.3\times 10^{-6}$ & $5.9\times 10^{-6}$ \\
0.3 & $1.6\times 10^{-6}$ & $1.2\times 10^{-6}$ & $5.4\times 10^{-6}$ \\
1 & $1.9\times 10^{-6}$ & $1.5\times 10^{-6}$ & $6.6\times 10^{-6}$ \\
3 & $3.2\times 10^{-6}$ & $2.3\times 10^{-6}$ & $1.1\times 10^{-5}$ \\
5 & $4.4\times 10^{-6}$ & $3.2\times 10^{-6}$ & $1.4\times 10^{-5}$ \\
10 & $7.3\times 10^{-6}$ & $5.4\times 10^{-6}$ & $2.4\times 10^{-5}$ \\
\hline
\end{tabular}
\end{center}
\caption{Bounds on the imaginary parts of the mass insertions for the parton 
neutron model. 
For   the squark masses different from 500 GeV, the bounds are to be multiplied
 by $m_{\tilde q}/500\; GeV$.}
\end{table}

\begin{table}
\begin{center}
\begin{tabular}{|c||c|c|c|}
\hline
  $x$   & $\vert {\rm Im}(\delta_{11}^{d})_{LR} )  \vert$ & $\vert {\rm Im}(\delta_{11}^{u})_{LR} )  \vert$ &
$\vert {\rm Im}(\delta_{22}^{d})_{LR} )  \vert$ \\
\hline
\hline
0.1 & $2.6\times 10^{-8}$ & $2.6\times 10^{-8}$ & $2.2\times 10^{-6}$ \\
0.3 & $3.6\times 10^{-8}$ & $3.6\times 10^{-8}$ & $3.0\times 10^{-6}$ \\
1 & $6.7\times 10^{-8}$ & $6.7\times 10^{-8}$ & $5.6\times 10^{-6}$ \\
3 & $1.5\times 10^{-7}$ & $1.5\times 10^{-7}$ & $1.3\times 10^{-5}$ \\
5 & $2.5\times 10^{-7}$ & $2.5\times 10^{-7}$ & $2.1\times 10^{-5}$ \\
10 & $5.3\times 10^{-7}$ & $5.3\times 10^{-7}$ & $4.4\times 10^{-5}$ \\
\hline
\end{tabular}
\end{center}
\caption{Bounds on the imaginary parts of the mass insertions 
imposed by  the mercury EDM.
For   the squark masses different from 500 GeV, the bounds are to be multiplied
 by $m_{\tilde q}/500\; GeV$.}
\end{table}

\newpage

\end{document}